\documentclass[prb,reprint,superscriptaddress,amsmath]{revtex4-2}
\usepackage[utf8]{inputenc}
\usepackage{graphicx}
\usepackage{bm}
\begin{document}

\title{On the Onset of Coherent Phonon Motion in Peierls-Distorted Antimony by Attosecond Transient Absorption}

\author{Lauren B Drescher}\email{lauren.drescher@mbi-berlin.de}
\affiliation{Department of Chemistry, University of California, Berkeley, California 94720, USA}
\affiliation{Chemical Sciences Division, Lawrence Berkeley National Laboratory, Berkeley, California 94720, USA}
\affiliation{Department of Physics, University of California, Berkeley, California 94720, USA}
\author{Bethany R. de Roulet}
\affiliation{Department of Chemistry, University of California, Berkeley, California 94720, USA}
\affiliation{Chemical Sciences Division, Lawrence Berkeley National Laboratory, Berkeley, California 94720, USA}
\author{Yoong Sheng Phang}\altaffiliation[Current Address: ]{Department of Physics, Harvard University, Cambridge, Massachusetts 02138, USA}
\affiliation{Department of Chemistry, University of California, Berkeley, California 94720, USA}
\affiliation{Department of Physics, University of California, Berkeley, California 94720, USA}
\author{Stephen R. Leone}
\affiliation{Department of Chemistry, University of California, Berkeley, California 94720, USA}
\affiliation{Chemical Sciences Division, Lawrence Berkeley National Laboratory, Berkeley, California 94720, USA}
\affiliation{Department of Physics, University of California, Berkeley, California 94720, USA}

\date{June 24, 2025}

\begin{abstract}
Attosecond extreme-ultraviolet (XUV) transient absorption spectroscopy measurements on the Peierls-distorted phase of the semimetal antimony (Sb) are presented.
After excitation by an ultrashort, broad band near-infrared (NIR) pulse, the distortion is (partly) lifted causing the well-known coherent phonon motion of the lattice. 
While the overall observed dynamics generally follow a displacive excitation model, a delayed onset of the pump-induced carrier dynamics due to hot-carrier thermalization is observed, as well as a large spectral phase dependence in the coherent phonon oscillation. 
This is attributed to significantly different carrier relaxation timescales for carrier energies above and near the Fermi level of the semimetal and corroborated by a simple theoretical model that considers the carrier relaxation timescales in the displacive phonon model to explain the observed dynamics. 
Our results provide direct experimental evidence about the role of carrier-relaxation in the origin of displacive coherent phonon motion.

\end{abstract}

\maketitle

\section*{Introduction}
The coupling of optical excitation to atomic forces and lattice deformation offers intriguing ways to reach non-equilibrium and meta-stable states, with interesting proposed applications including electronic\cite{caviglia2012, horstmann2020} and magnetic switching\cite{afanasiev2021a, davies2024, stupakiewicz2021}. 

The use of coherent optical pulses allows these transformations to occur on ultrafast timescales and with potentially higher efficiency compared to incoherent thermal pathways~\cite{delatorre2021}. 
To optimally develop and engineer these non-thermal pathways towards out-of-equilibrium phases, precise understanding of the underlying mechanisms in which the optical excitation couples to phonon motion needs to be developed.

While coherent phonons can be created directly out of coupling to the electromagnetic radiation, e.g. from THz frequency photons or Raman excitation by higher frequencies, an indirect mechanism termed “displacive excitation of coherent phonons” (DECP) is known to be prominent~\cite{zeiger1992} and was shown as a promising route to control phonon motion~\cite{hase1996}. 
In this mechanism the photon-driven electronic excitation changes the energetic landscape of the solid, leading the lattice to rearrange to a new energetic optimum position, launching an ultrafast coherent optical phonon motion. 

The effect is exemplary in Peierls-distorted materials: Peierls-distortion describes a one-dimensional lattice deformation at equilibrium, in which the atomic sites in the lattice are distorted from the equally-spaced pure-lattice energetic minimum, and the less symmetric configuration allows the electrons to occupy lower energy bands than in the more-symmetric lattice, achieving an overall energetically favorable configuration. 
If, however, these electrons are forced out of their equilibrium energy due to the optical excitation, the lattice is now free to lift the deformation and moves towards the higher symmetry phase. 

In their original work on DECP, Zeiger \textit{et al.}~\cite{zeiger1992} describe a simple model of this motion and successfully explain the observed temporal behavior of the NIR reflectivity due to atomic forces acting on the lattice following optical excitation. 
Early follow up work identified a deviation in the phase of the phonon motion in antimony (Sb) that could not be explained by this displacive model~\cite{garrett1996}, showing limitations in the understanding of the creation of this motion.
Recent experimental and theoretical work has identified additional forces acting on the lattice during the relaxation of optically excited carriers as a potential source of this phase~\cite{li2013,omahony2019}, i.e. the excited carriers exact a different force on the lattice directly following optical excitation compared to when they have reached a thermal distribution.
The theoretical results implied that these additional forces could affect the efficiency of the conversion of optical excitation into atomic motion, which would be desired for the proposed applications.
To address measurements of such a phase, extremely short timescales are required.
Recent experiments using time-resolved spectroscopy with temporal resolutions on the order of a few femtoseconds (fs) of metals~\cite{chang2021a}, semiconductors~\cite{zurch2017,cushing2018,attar2020} and semimetallic graphite~\cite{bandaranayake2020, rohde2018}, have shown that there can be significant observable differences between the creation of excited carriers by a pump pulse and the formation of a thermalized distribution through carrier-carrier and carrier-phonon scattering (carrier thermalization).
In contrast in the derivation of the model equations, Zeiger \textit{et al.} note that both the number of optically (i.e. non-thermal) excited electrons $\langle n \rangle$ and the change of electron temperature $\Delta T_e$ lend themselves as the basis for their model and that they expected equal outcomes for both. 
This leads to the opportunity to explore whether carrier relaxation affects the coherent phonon motion.
In this work, we explore the behavior of the coherent phonon signal following displacive excitation by an NIR pulse of Peierls-distorted Sb using attosecond XUV transient absorption spectroscopy, which has been shown to be sensitive to coherent phonon motion through energetic shifts of the observed absorption~\cite{weisshaupt2017,geneaux2021b}.

While it is well known that the relaxation of carriers in solids are generally many-body functions, i.e. they not only depend on their momentum and band index or energy, but also that of all other carriers, several approximations can often be made to simplify their description~\cite{rossi1994}. While a two-temperature model simply assumes the distribution of carriers to always represent a thermal (i.e. Fermi-Dirac) distribution and interacts thermally with the phonon population, this assumption necessarily breaks down at early-enough time-scales after photo-excitation where carriers might be energetically distributed far out of their thermal equilibrium. A first step to regain the complexity of the underlying dynamics is to extend the TTM by introducing a population of non-thermal carriers that decays with a constant decay rate, which we will call the extended TTM here. Further inclusion of the energy and momentum dependent relaxation rates, state- and population density leads to the popular Boltzmann equation model.

As we will see in this work, while the TTM well describes the transient XUV absorption dynamics of Sb, at early timescales it indeed brakes down and requires the introduction of the extended TTM. We will further see, that while the spectrally dependent carrier relaxation rates are not directly observable within the carrier population signal within the signal-to-noise rate of our experiment, they still observably influence the phonon motion.

This work is structured as follows: After a brief introduction of the sample, results on the static and transient XUV absorption of Sb are presented. The results are first analyzed by a component analysis. The components show that the dynamics predominantly follow a two-temperature model that allows to extract the (sub)ps scale dynamics of phonon oscillation and decay, carrier cooling and lattice heating. Next, the disparities between the model and the measured spectra are evaluated on the fs timescale following optical excitation. The effects of non-equilibrium optically excited carriers and the breakdown of the two-temperature model at early timescales are highlighted, and spectrally dependent carrier thermalization and cooling rates are obtained. Then, the phase of the coherent phonon oscillation is analyzed and the effect of carrier-relaxation on the phonon motion is discussed. 
Finally, differences in the dynamics at the onset of the transient absorption during the optical excitation are discussed.

The results provide experimental proof that the response of the coherent phonon motion is indeed affected by the relaxation of optically-excited carriers into the thermalized, hot Fermi-Dirac distribution, both in the amplitude of the observed signal as well as in the phase of the phonon oscillation.

This work significantly extends a previous study of the XUV transient reflectivity spectroscopy of the coherent phonon motion of Peierls-distorted Bi~\cite{geneaux2021b}. The component analysis presented here gives equivalent results compared to a spectral modelling fitting as introduced in previous XUV transient spectroscopy studies~\cite{geneaux2021b,zurch2017,cushing2018,sidiropoulos2023}. The increased experimental sensitivity and time-delay sampling in this study allows the detailed exploration of the early-timescale dynamics, elucidating the role of photoexcitation and carrier relaxation on the phonon motion and the Peierls distortion.

\section*{Results}
In Sb, the Peierls distortion leads to a displacement of neighboring hexagonal planes along the c lattice coordinate, with the so-called Peierls parameter $z=0.2338$ at 298\,K (whereas $z=0.25$ would correspond to equidistant spacing of hexagonal planes and a cubic lattice symmetry)~\cite{barrett1963}.
The distortion leads to an opening of a pseudo-gap at the Fermi energy ($E_F$) and the Sb semimetallicity, which exhibits a narrow density of states (DOS) around $E_F$ - characteristic for the group V semimetals~\cite{gonze1990}. 
In addition to the original DECP studies on Bi and Sb, coherent phonon motion is well studied by transient reflectivity and absorption in the NIR to visible~\cite{zeiger1992,garrett1996,ishioka2008,hase2015,teitelbaum2018a} and XUV spectral regions~\cite{geneaux2021b}, as well as photoelectron spectroscopy~\cite{papalazarou2012} and time-resolved diffraction~\cite{fritz2007,waldecker2017,teitelbaum2018}.
Recent theoretical studies using density-functional theory~\cite{giret2011,murray2015,omahony2019} (DFT) and molecular dynamics~\cite{bauerhenne2017} (MD) simulations have successfully treated the excitation of the coherent phonon motion due to a change of the minimum coordinate of the potential energy surface as a function of electron temperature or number of excited electrons.

A band-structure diagram and (projected) density of states plot (DOS) of Sb at equilibrium, obtained via DFT (see Appendix~\ref{app:dft}), can be seen in Fig.~\ref{fig:static}e. Typical for the semimetallic properties, the density of states is minimal around the Fermi level with small pockets below the Fermi level around the L-point. The bands mostly exhibit $p$-orbital character, indicating excellent spectroscopic visibility from the $4d$ core levels.
For the spectroscopic experiments, polycrystalline thin film samples of Sb of 32\,nm thickness on Si$_3$N$_4$ membranes are prepared (see Appendix~\ref{app:sample}).
The laser pulses used for the experiments are 4\,fs duration near infrared pulses from a Ti:sapphire laser for excitation (see Fig.~\ref{fig:static}b) and a short pulse train of attosecond pulses in the extreme ultraviolet for the probe (see Fig.~\ref{fig:static}a), produced by the process of high harmonic generation (see Appendix~\ref{app:setup}).

\subsection{Static Absorption}
\begin{figure*}[tb]
\includegraphics[width=\textwidth]{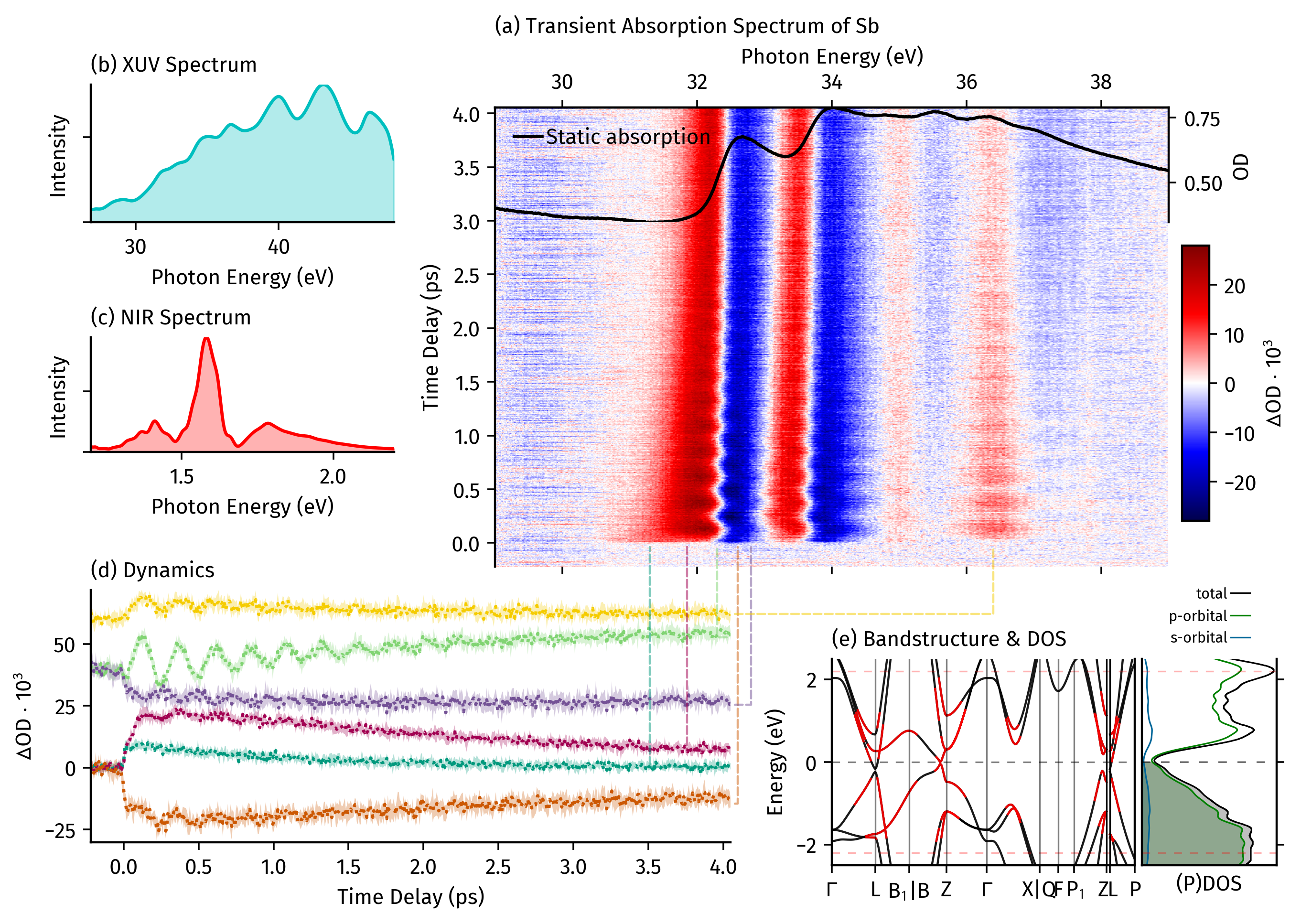}
\caption{\textbf{a} Transient absorption spectrum of Sb in the region between 29\, and 39\,eV. After the initial excitation the change in absorbance is modulated by the coherent phonon motion that quickly decays within the first picoseconds. \textbf{Inset} Static absorption spectrum of Sb in the vicinity of the 4d core level transitions (cf Fig.~\ref{fig:components}). \textbf{b} XUV Spectrum and \textbf{c} NIR Spectrum used in the experiment. \textbf{d} Transient dynamics at selected energies. The dashed lines indicate the spectral photon energy in (a) at which the dynamics were taken. Purple and teal lines are representative for the dynamics of holes in the valence bands, while orange and violet lines are representative for the dynamics of electrons in the conduction bands. The light green and yellow lines show dynamics dominated by phonon motion of the lattice, leading to a shifting of the core-to-Fermi and transitions from the core to higher CBs, respectively. Dynamics are vertically offset for clarity. The shaded areas around points indicate the CI from consecutive measurements (see Appendix~\ref{app:CI}). \textbf{e} Left: Band-Structure diagram of Sb at equilibrium geometry, energy relative to the Fermi level. Bands that energetically allow for direct, one-photon resonant transitions from valence to conduction band by the pump pulse shown in (c) are colored in red. Right: (Projected) Density of states, showing the typical minimum in density at the Fermi level. 
.\label{fig:static}}
\end{figure*}

In the inset of Fig.~\ref{fig:static}c, the static absorbance of Sb in the region between 29\,eV and 39\,eV is shown.
Dominating are two peaks at 32.6\,eV and 33.9\,eV, corresponding to transitions of electrons from the 4d$_{5/2}$ and 4d$_{3/2}$ (semi-)core levels of Sb in the conduction bands (CB). 
Towards lower photon energies a falling background absorption can be observed that is associated with high-energy valence transitions (i.e. not core-level transitions).
The absorption onset leads into a plateau region of absorption with additional peaks at 35\,eV and 36\,eV.
The structures agree well with previous experiments on the XUV absorption spectrum of Sb~\cite{ejiri1978}, from which we accordingly assign the core-to-Fermi-level transition energies at 32.1\,eV and 33.4\,eV for the 4d$_{5/2}$ and 4d$_{3/2}$ levels, respectively.
Using spin-orbit decomposition (see Appendix~\ref{app:SO}), the absorption from the $j=5/2$ core-level is reconstructed (see Fig.~\ref{fig:components}a). By comparing to the DOS (cf. Fig.~\ref{fig:static}e),  we can associate these peaks with high-symmetry points in the conduction bands, such as around the $\Gamma$ point.

\subsection{XUV Transient Absorption of S\lowercase{b}}
Next, the long-time-range transient absorption of Sb is taken. Spectra are recorded for time delays between –0.2\,ps and 4.0\,ps in 6.7\,fs steps, with an estimated incident fluence of 4.4\,mJ/cm$^2$, leading to an estimated photoexcited carrier density of $1.5\cdot10^{21}$\,cm$^-3$, or approximately 1\% of valence electrons. The fluence was chosen to obtain a high signal-to-noise ratio while staying sufficiently below sample degradation. The Gaussian instrument response function (IRF), dominated by the duration of the pump pulse, is determined to be $5.4\pm0.1$\,fs (FWHM, see Appendix~\ref{app:timezero}).
A small residual signal is observed at delay times where the NIR pulse arrives long after the XUV pulse. 
From our previous knowledge of measurements on thin film samples, this is typically due to a laser-induced heating.
This small contribution is subtracted equally from all spectra. 
The resulting delay-dependent change in absorbance is shown in Fig.~\ref{fig:static}c (for a complete description of data acquisition and analysis, see Appendix~\ref{app:datapipeline}).
At delay settings where the XUV probe precedes the NIR pump, no significant change in absorbance is visible. 
At the zero-time delay, a strong change in the signal is observed. 
An increase in absorbance occurs at energies below the onset of the 4d to CB absorbance of the static spectrum and a decrease in absorbance at energies above. 
The feature is repeated for both the spin-orbit split 4d$_{5/2}$ and 4d$_{3/2}$ orbitals. 
Furthermore, changes in absorbance are observed both within the energy range around $E_F$ accessible by the broad band NIR pump (e.g. roughly below 36\,eV, see shaded area in Fig.~\ref{fig:static}(e) right), and above. 
This shows that the observed dynamics are both due to a change in allowed core-level transitions following a pump-induced change in band-occupations (state opening/blocking) and overall changes in the electronic band structure.
At early timescales the increased absorption features extend to lower energies compared to later timescales, indicating a rapid rearrangement of the carriers within the first hundreds of femtoseconds after excitation. 
Additionally, the induced changes slowly decay towards longer time delays. 
Most strikingly, as observed in other investigations~\cite{weisshaupt2017,geneaux2021b,sidiropoulos2021,kato2020}, the change in absorbance exhibits strong oscillatory features at positive delay times. 
These oscillations appear to modulate the observed changes in absorption, which quickly decay towards greater time delays, up to around 1.0 to 1.5\,ps, i.e. faster than the overall decay of the pump-induced changes in absorption. 
These can be associated with the coherent phonon motion.  

\subsubsection{Component Analysis}\label{sec:components}
\begin{figure}[tb]
\includegraphics[width=0.5\textwidth]{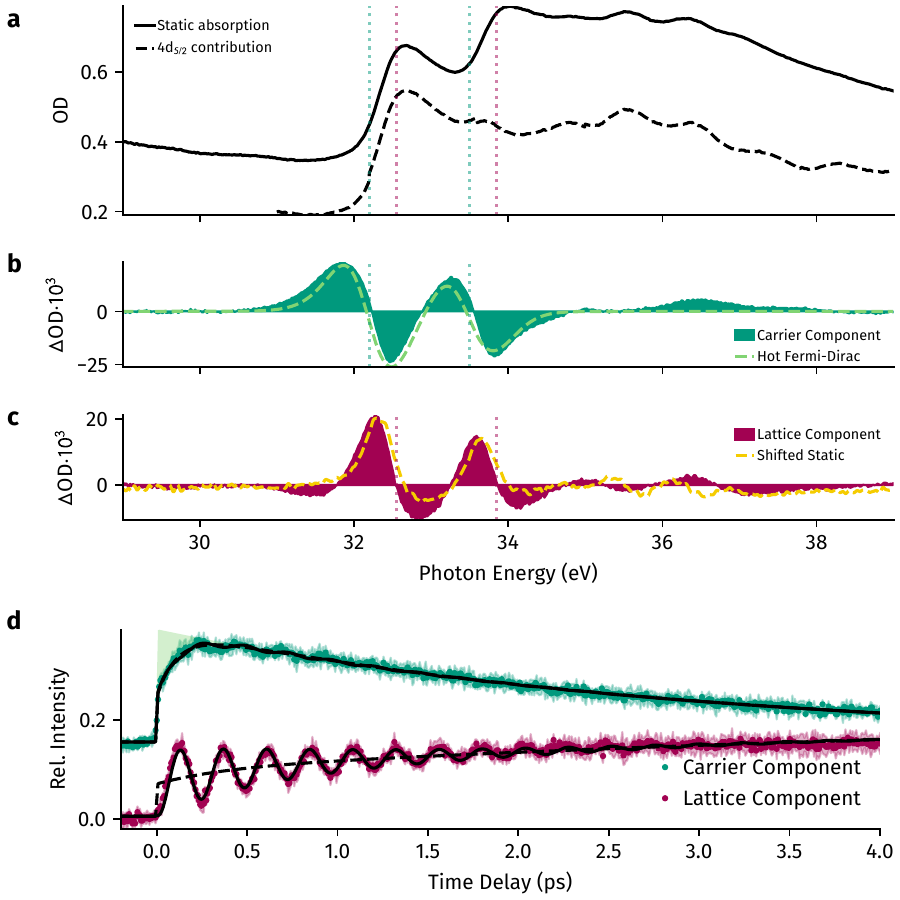}
\caption{\textbf{a} Static absorption spectrum of Sb in the vicinity of the 4d core level transitions. The peaks (dashed orange) at 32.6\,eV (4d$_{5/2}$) and 33.9\,eV (4d$_{3/2}$) are associated with transitions at high symmetry points (mostly L, B and Z) of the conduction band. The Fermi-energy~\cite{ejiri1978} of the transitions from the 4d$_{5/2}$ and 4d$_{3/2}$ core levels are indicated by the dashed blue lines. Deconvolution of the 4d$_{5/2}$ core level contribution (dashed black) show that structures in the static absorption at higher energies are attributed to transitions into higher conduction bands such as around the $\Gamma$ point. \textbf{b} Change in absorbance due to increase of carrier temperature and \textbf{c} due to lattice motion, obtained by component analysis (blue and orange, respectively). The carrier temperature component can be compared to two hot Fermi-Dirac distributions around the core-to-Fermi transition energies (dashed pink) while the lattice component fits well to the difference in absorbance due to a shifted static absorption (dashed brown). \textbf{d} Temporal dynamic of the singular value components (teal and purple dots), showing the rapid onset and subsequent decay, as well as oscillations caused by coherent phonon motion. The best-result fit (see Appendix~\ref{app:model}) is shown (black solid lines), as well as the modeled signal without phonon motion (dashed lines, see text). Notably, the hot carrier signal does not rise with the pump-pulse, but the maximum appears delayed (teal shaded area). Teal dots and associated lines are offset from zero for clarity. Confidence intervals of the components are estimated from consecutive measurements (see Appendix~\ref{app:CI}).\label{fig:components}}
\end{figure}

To allow further insight into the observed dynamics and to reduce the complexity of the observed signal, a component analysis approach via singular value decomposition (SVD) is chosen. A short introduction into the method is given in Appendix~\ref{app:svd}. The SVD yields ``global'' spectral and temporal components from the experimental data, i.e. without the need to further select certain temporal or spectral ranges and without fitting of an underlying physical model.
Here this global analysis is performed over the full spectral and delay range shown in Fig.~\ref{fig:static}c.
To perform the SVD in the basis of the static absorbance, the static absorbance is added back to the measured change in absorbance. The leading component, encoding the static absorbance, is then omitted from the following analysis.
The resulting spectral components (Fig.~\ref{fig:components}b,c) and temporal components (Fig.~\ref{fig:components}d) for the next two strongest components are resolved (a discussion of the residual components is presented in the SI): 
Analyzing the change in absorbance of the spectral components, we note that the first component (Fig.~\ref{fig:components}b) shows a strong, asymmetric lineshape around the Fermi-energy of the core-level transitions (32.1\,eV and 33.4\,eV). This is consistent with a broadening of the Fermi-Dirac distribution due to an increase in carrier temperature. Indeed, fitting a Fermi-Dirac distribution (convoluted with the DOS) for an increase in peak carrier temperature of ~2000\,K to both spin-orbit core-to-Fermi transitions yields good agreement with the obtained distribution and the thermal energy of the absorbed pump pulse. On the other hand, the second component (Fig.~\ref{fig:components}c) shows a similar asymmetric lineshape around the energy of the absorption peaks. This is consistent with an overall shift of the absorption spectrum due to movement of the lattice, which has been found to be the predominant effect of the coherent phonon motion~\cite{geneaux2021b}. This can be confirmed by comparing the component to the change in absorbance that a shifted static absorption would yield. Good agreement is found for a shift of the static of -28\,meV at the largest excursion.  Therefore, the first component is associated with the change in carrier temperature and the second component with the dynamic of the lattice motion. 
Small deviations of this can be observed: In case of the lattice signal, while recent DFT calculations for Bi~\cite{geneaux2021b} have shown that the change of the Peierls coordinate leads to an energy shift of the absorption. This is caused by a closing of the negative band gap and a renormalization of the conduction bands closer to the Fermi level, leading to an overall red-shift of the core-to-CB absorption. At the same time, the effect is counter-acted by the reduced Coulomb repulsion between the ions, which leads to a higher binding energy of the core-levels, which leads to a (smaller) blue-shift of the core-to-CB transition~\cite{geneaux2021b}. While the change in absorption can be approximated well by comparison the shifted static absorption, the component analysis applied here could potentially be better suited to represent the more detailed changes due to renormalization of the bands compared to a fitted modelling approach, which considers only changes from a shifted absorption. While the energy of the observed red-shift of the absorption due to the lattice motion is similar to the quasi-particle energy associated with the phonon oscillation, the two are (within first order) not connected, since the red-shift is dependent on the oscillation amplitude and not the frequency. 
In the carrier signal, differences between the hot Fermi-Dirac distribution and the observed spectrum could be due to different transition strengths between populations in different bands. At higher energies, a broad feature around 36.5\,eV is observed in the carrier component. These photon energies are above the resonant one-photon excitation window of the pump pulse from the VB. It is unlikely that  they originate from a non-linear or multiphoton excitation process, as an occupation of states high in the CB would be expected to yield a negative change of absorbance, compared to the observed increase. A possible origin of this feature is a change of band structure in the higher valence bands due to photodoping and many-body effects by the large number of excited carriers~\cite{feneberg2014}.

The observation seems to indicate that the component analysis follows a two temperature model (TTM), where a laser-induced heating of the carrier temperature exchanges energy with the lattice subsystem.
This is furthermore supported by the temporal dynamics:
We observe a strong rise and subsequent exponential decay following zero-delay in the first component (teal dots in Fig.~\ref{fig:components}d) and a strong, exponentially damped oscillatory modulation and a slow rise in the second component (purple dots).
The fast rise and exponential decay of the first component is characteristic for an effect of a change in carrier temperature, while the oscillatory behavior and the slow rise of the second component can be associated with a change of lattice structure due to the phonon motion and the heating of the lattice.
The carrier temperature signal appears to be modulated by the phonon oscillation, and the maxima in the carrier temperature coincide with minima in the lattice signal and vice-versa.
A similar effect had been previously theoretically predicted~\cite{giret2011} and was observed in the XUV transient reflectivity spectroscopy of bismuth~\cite{geneaux2021}, where the result was associated with a modulation of the signal due to the back-coupling interaction of the lattice shift by the coherent phonon motion on the specific heat capacity of the carriers and therefore their temperature. 

This shows that a component analysis can be highly effective in analyzing femto- to picosecond dynamics of XUV transient absorption spectra, without the need for prior physical modelling and information from (time-dependent) DFT, as would be required in spectral fitting approaches.

Notably, however, and in conflict with a two-temperature model, the signal associated with a change in carrier temperature does not immediately reach its maximum within the instrument response function (IRF), but appears delayed (see shaded area in Fig.~\ref{fig:components}d).
While a delayed build-up dynamic, as observed in the carrier-signal here, has been observed in reflectivity measurements in Bi and Sb before, it has been largely associated with a delayed response of the electronic system, but has found little discussion.  
Similar effects have been observed for XUV transient absorption spectroscopy in semiconductors~\cite{zurch2017,cushing2018,attar2020}, where they were related to relaxation of optically excited carriers to a thermodynamic equilibrium via intra-valley carrier-carrier scattering or inter-valley carrier-phonon scattering. 
On the other hand, Giret~\textit{et al.}~\cite{giret2011} have discussed effects of carrier diffusion along the layer thickness in X-ray diffraction experiments~\cite{fritz2007}. Extending a two-temperature model to include one-dimensional carrier diffusion to explain the delay, we are unable to find a sufficient diffusion effect on the carrier temperature, even when varying reported diffusivity values for Sb over orders of magnitudes.
This is also supported by evaluating the residual components, which show a large signal in the first ~200\,fs following photoexcitation, implicating that the carrier occupation does not follow a thermal distribution at these early timescales (Residuals are shown in Appendix~\ref{app:residuals} and the carrier occupation following photoexcitation is discussed in more detail below).
This shows, that the TTM is insufficient to describe the carriers during the first few femtoseconds following optical excitation, during which the carriers occupy highly excited states that do not follow a thermal distribution and are not in thermodynamic equilibrium with each other. 

While more complex modelling based on time-dependent Boltzmann equations would be able to more accurately describe the band-dependent occupations during thermalization, we find that an extended two temperature model with a classical treatment of the phonon motion that treats the non-thermal carriers as an effective population (see Appendix~\ref{app:model}) sufficiently covers the observations and allows for a meaningful quantization of the observed dynamics within the component analysis.
Within this model, the non-thermal carriers relax into the thermal pool of carriers and increase the effective temperature of the thermalized carriers. 
The best-fit result of a non-linear least-squares fit of the model to the carrier temperature and lattice dynamic is shown in Fig.~\ref{fig:components}d (black solid lines). All obtained values are shown in the appendix (Table~\ref{tab:fit}) and the time-dependent properties are discussed below.

\subsubsection{Interpretation of Fitted Model Dynamics}
Compared to previous results of the Sb coherent phonon mode, measured by transient reflectivity studies in the NIR spectrum (obtaining 4.5 THz)~\cite{zeiger1992}, a smaller value of the A\textsubscript{1g} phonon frequency of 4.19$\pm$0.01 THz is observed from the fitting for these particular thin film samples.
Similarly, a longer lifetime for the carrier recombination (2.95$\pm$0.04\,ps) and a much shorter lifetime of the coherent phonon motion (0.96$\pm$0.04\,ps) is obtained, compared to reference data (1.67\,ps and 2.90\,ps, respectively)~\cite{zeiger1992}. 
There are several possible explanations for these differences. 
First, as the observation of a residual heat signal shows, the very-thin samples are heated by the laser excitation to temperatures above room temperature. 
It was shown that higher temperatures lead to a softening of the phonon frequency, as well as a faster decay of the coherent phonon in Sb~\cite{hase2015}, as well as in Bi~\cite{boschetto2008} and Ti$_2$O$_3$~\cite{zeiger1996}. 
However, in the case of Ti$_2$O$_3$ the increase in temperature was also associated with a fast recombination of carriers, which is opposite to the elongated lifetime of carriers observed here. 
Secondly, a similar effect on the softening and phonon lifetime has been observed to be dependent on the pump fluence for Sb~\cite{omahony2020}, Bi~\cite{decamp2001} and Te~\cite{hunsche1995}. 
Lastly, the (polycrystalline) thin films are known to influence both the lifetimes of photoexcited carriers as well as the coherent phonon mode, due to their higher defect density compared to single crystals and also due to confinement within the films.
While we do not observe trapping of photoexcited carriers due to defect states in the experiment directly, the observed timescales of the carrier dynamics are likely to be affected by the expected large defect density of the thin films. A direct comparison between a single crystalline sample and a thin-film polycrystalline sample could further elucidate this connection.
For the coherent phonon frequency, effects due to defect density and confinement would lead to a shift in frequency for the A\textsubscript{1g} mode in the Raman spectrum as well, which was found to be close to literature reference values (4.4\,THz in the Raman spectrum here vs. 4.5\,THz reported~\cite{zeiger1992}). 
Extrapolating changes in the phonon decay rate in Sb that have been observed for temperatures below 300\,K to higher temperatures below the melting point of Sb would not yield the observed results~\cite{hase2015}. 
Comparison to measured fluence dependencies of the phonon frequency and decay rate by NIR single shot transient absorption spectroscopy on Sb thin films~\cite{omahony2020} yields good agreement with the values obtained here. 

The phase of the phonon oscillation from the fitting procedure (-0.09±0.02\,$\pi$rad) clearly differs from the pure DECP model, which only predicts a phase-effect from the damping (which we find accounts to $<-0.01$\,$\pi$rad)~\cite{zeiger1992}. This indicates a delayed reaction of the phonon motion. Reported values on the A\textsubscript{1g} phase in the literature are varying, ranging from -0.02$\pm$0.02\,$\pi$rad~\cite{zeiger1992} to -0.13$\pm$0.01\,$\pi$\,rad ~\cite{garrett1996} and warrant a closer inspection later.

We do find good agreement between the modeled delayed rise of the carrier temperature due to carrier thermalization to the experimental observation, however it is necessary to include an additional scaling factor $r=0.60\pm0.03$ that limits the effect of the thermalization to achieve a good fit.
This scaling factor describes that only about 60\% of the signal associated with the hot carriers shows an observable relaxation timescale, while the remaining 40\% seem to appear within the instrument response function (IRF).
This implies either part of the excited carrier population relaxes into the thermalized carrier distribution within the duration of the pump pulse, while other carriers thermalize on measurably longer timescales, or a part of the excited carrier population already occupies the same states that are within their thermal distribution.
Recent relaxation rate calculations by O'Mahony~\cite{omahony2020} show a long lifetime for carriers at or below $E_F$, comparable to the result of the fit ($117\pm9\,fs$), but predict fast relaxation ($<5$\,fs) for carrier energies $>0.5$\,eV above or below $E_F$. This is due to the larger density of states in those regions, leading to a much faster carrier relaxation than in the narrow regions closer to $E_F$.
The observed relaxation timescale ($117\pm9\,fs$) is furthermore comparable to previous observations in NIR reflectivity~\cite{ishioka2008}.
Here we use the word thermalization to refer to the possibility of several processes, carrier-carrier scattering and carrier-phonon scattering, reaching a hot Fermi Dirac distribution and these can be carrier-energy dependent.
While relaxations via carrier-carrier scattering have reported lifetimes of a few femtoseconds, relaxation via carrier-phonon scattering has been reported to occur on tens to hundreds of femtoseconds, making it the likely dominating observed mechanism.

\subsection{Spectral Dependent Dynamics}
\begin{figure}[tb]
\includegraphics[width=.5\textwidth]{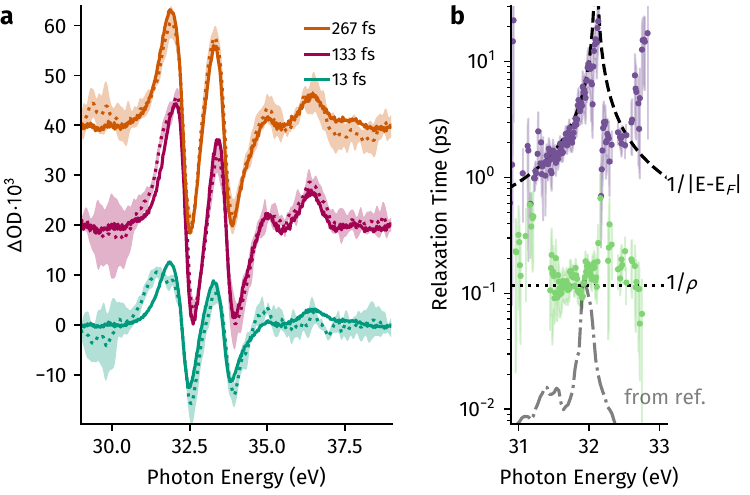}
\caption{\textbf{Spectral Dependence of Carrier Relaxation.} \textbf{a} Comparison between the predicted change in absorbance from the TTM fit and lineouts of the experimental data at several delays. The strong differences from optically excited carrier occupation, shortly after the pump pulse is over, quickly dissolve as the carriers relax into a thermal distribution. Spectra are vertically offset from zero for clarity. Confidence intervals of the experimental data are estimated from standard deviation between consecutive measurements. \textbf{b} Energy-dependent occupation lifetimes from biexponential fits. Two lifetimes are necessary to describe the observed dynamics, a long decay (violet) which scales to the occupation from a thermal distribution (dashed line) and a short rise or decay for the initial redistribution (thermalization) from optical excitation into thermal distribution (light green), centered around the lifetime obtained from the component analysis, $1/\rho=117\pm9$\,fs (dotted). The thermalization timescale appears flat within the confidence interval (variance of the non-linear least squares fit) and only agrees with predicted state lifetimes~\cite{omahony2020} (dotted-dashed line) within a narrow region.\label{fig:residual}}
\end{figure}

To examine the break-down of the two temperature model after the initial excitation by the pump pulse and the following carrier thermalization more closely, the measured transient absorption spectrum at early time delays are compared to the predicted spectrum from the TTM. As can be seen in Fig.~\ref{fig:residual}a, the experimental data show discrepancies to the model due to the large number of non-thermalized carriers shortly after the pump pulse is over (12\,fs). Compared to the model, a larger absorbance is found in a region ~1\,eV below the Fermi level, indicating that the pump excites carriers largely from the lower bands at the L- and Z-point (cf. Fig.~\ref{fig:static}e). On the other hand, at energies up to 0.5\,eV below the Fermi level, there is less absorbance than expected from the model, indicating that fewer holes are initially in the highest valence bands. Above the Fermi level, a stronger negative change in absorbance is observed, peaking at 32.6\,eV (i.e. 0.5\,eV above $E_F$), indicating a stronger population of the conduction bands at the L- and Z-point than in the thermal model. The effect is weakened due to the overlap of the conduction band signal from the spin-orbit split core-level.

This difference between the initial optical excitation and the thermalized carrier occupation can be seen in the dynamics as well: To confirm the behavior obtained from the TTM fitting, the rates from a bi-exponential rise and decay function are fitted to the transient absorption spectrum: A fast rise/decay to account for the initial thermalization of carriers, a slow decay for the carrier cooling and a damped oscillator term to account for the phonon motion (for details see Appendix~\ref{app:biexponential}). To suppress the influence of the upper spin-orbit split core transition from the fit, a spin-orbit decomposition is applied first (see Appendix~\ref{app:SO}). Only results where the fit successfully converges to physical measurements are reported. The resulting lifetimes are shown in Fig.~\ref{fig:residual}b. On the picosecond scale, the signal relaxes with a rate $1/\tau\propto |E-E_F|$ (dashed line). This is due to the exponential cooling of the electron temperature that changes the energy dependent band-occupation in the Maxwell-Boltzmann limit of the Fermi-Dirac distribution ($n_\textrm{Thermal}\propto\exp(-|E-E_F|/k_BT)$), confirming the thermal character of the signal at long timescales. Recently, Sidiropoulos \textit{et al.} have connected this energy-decay scaling to a limited dimensionality of the carrier system~\cite{sidiropoulos2023}.
For the fast timescale, little significant spectral dependence on the thermalization rate is found. It is close to the timescale found in the component analysis (dotted line in Fig.~\ref{fig:residual}b). In the valence bands, the lifetime seems to increase for energies further away from the Fermi level, although the decreased signal strength and therefore higher variance in the fit-result in this region prevents a detailed analysis. It should be noted, that the fast timescales represent a decay in the region below 31.6\,eV and a rise in the region above, as evident in Fig.~\ref{fig:residual}a. In the conduction band the agreement for both the slow and fast dynamics are worse, and here the fit convergence is poorer, since the transient absorption is strongly affected by the lattice dynamic.
Notably, the fitted lifetimes do not correspond to previously reported lifetimes of the electronic states in Sb from DFT calculations~\cite{omahony2020}, except for a small region close to the Fermi level that also corresponds to the largest reported lifetimes.
This implies that the thermalization of the initially optically excited states in the experiment was not limited by the state lifetime, but appears bottlenecked, either through limited final state density at the top of the valence band or lack of open channels to dissipate excess energy through carrier-carrier or carrier-phonon scattering.

\subsection{Spectral Phase of Phonon Motion}
\begin{figure*}[tb]
\includegraphics[width=\textwidth]{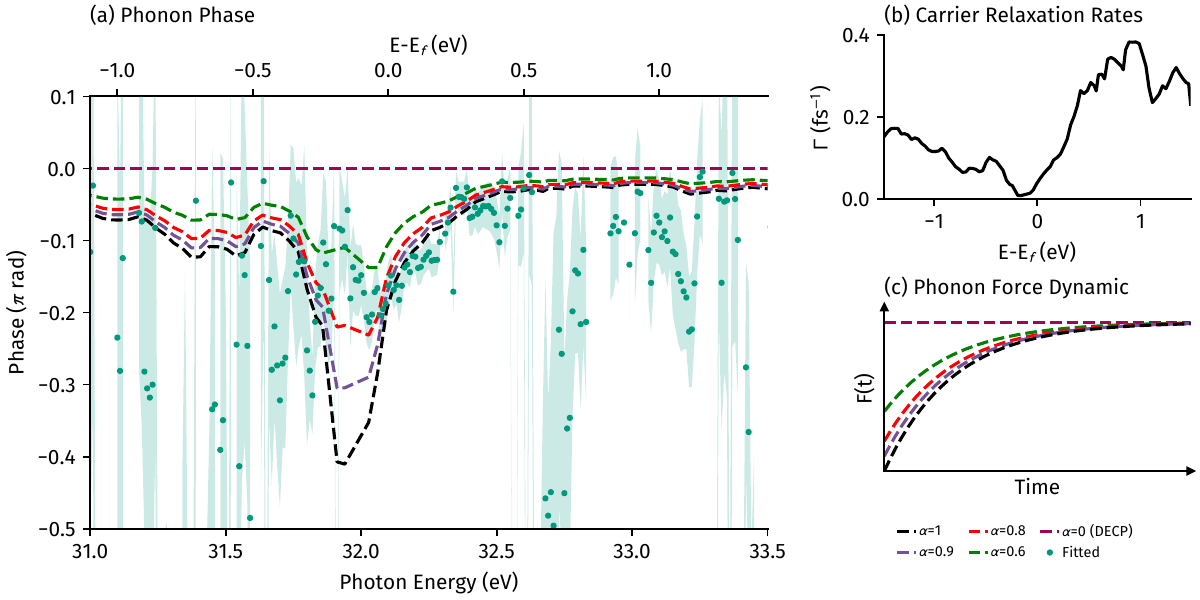}
\caption{\textbf{Spectral Phase of Phonon Motion.} \textbf{a} Phase of phonon motion as obtained through non-linear fit. The phase shows a minimum at the Fermi level (32.1\,eV). Comparison to phases from model equations (Eqn.~\ref{eq:model-solution}) using calculated relaxation rates (\textbf{b}, reproduced from \cite{omahony2020}) indicates this behavior is due to the interatomic force $F(t)$ rising from a fractional value (expressed by $\alpha$) during the carrier relaxation. \textbf{c} Illustration of the exponential rise of $F(t)$ (see Eqn~(\ref{eq:model-pdq})).
\label{fig:phase}}
\end{figure*}

In contrast to the relatively flat behavior of the thermalization timescale, a strong spectral dependence of the phase of the phonon motion is observed and shown in Fig.~\ref{fig:phase}a. In the region close to the Fermi level a sharp dip in the phase is observed with the minimum occurring at the Fermi level.
A second minimum can be seen slightly below the Fermi level, at around -0.3\,eV. These extrema coincide with extrema in the band-structure of Sb (cf. Fig.~\ref{fig:static}e).
At lower energies, the fitted phase widely diverges due to the limited spectral range of the phonon motion.
It is noted that while the phase shown here is obtained through a non-linear fit, compatible results can be obtained through Fourier analysis, as shown in Appendix~\ref{app:fourier}.

The phase difference of the phonon oscillation from a pure cosine form ($\varphi = n\pi$) has been associated with a mixing of the DECP mechanism with an impulsive excitation~\cite{garrett1996}.
In the case of Bi and Sb, O’Mahony~\textit{et al.} have calculated the interatomic force for different excited states~\cite{omahony2019}. For the A\textsubscript{1g} phonon mode they find an initially higher force caused by the carrier population after optical excitation at 1.55\,eV compared to the force after relaxation. This transient force adds an impulsive component to the displacive motion, which imparts an additional phase on the A\textsubscript{1g} mode and a reduction of the oscillation amplitude. 
However, such additional forces would lead to a phase of opposite sign compared to the measured phase here. It was suggested that the transient interatomic force of the optically excited carrier population could also be lower than after relaxation, e.g. in the case of As or for different photon energies in the case of Bi and Sb~\cite{omahony2019}.

In the following, the approach of O'Mahony~\textit{et al.} is adapted to study the effect of a time-dependent phonon force $F(t)$ that arises due to the displacive force $F_0$ with the rate of carrier relaxation $\rho$, with a parameter $\alpha$ that allows to test the influence of carrier-relaxation on the phonon force:
\begin{equation}
F(t) = F_0(1-\alpha e^{-\rho t}),\label{eq:model-pdq}.
\end{equation}
Within this model, $\alpha\to1$ corresponds to a limit in which only the relaxed carriers contribute to the phonon motion, i.e. at $t=0$ there is no force acting on the phonon. In contrast, $\alpha=0$ corresponds to all carriers contributing equally, i.e. the number of excited carriers $\langle n\rangle$ is driving the phonon motion and the constant displacive force is already established at the end of the optical excitation (see illustration in Fig.~\ref{fig:phase}c).

Inserting the time-dependent force in the phonon-coordinate $Q(t)$ equation of motion (cf. Eqn.~(\ref{eq:eom-phonon}) below), the solution to the differential equation:
\begin{equation}
Q(t) = A\cos(\omega t+\varphi)+\frac{1}{\omega^2}-\frac{\alpha e^{-\rho t}}{\rho^2+\omega^2},\label{eq:model}
\end{equation}
then describes a cosine-oscillating phonon motion with phase
\begin{equation}
\tan(\varphi)= -\frac{\alpha\rho\omega}{\rho^2-(\alpha-1)\omega^2} \stackrel{\alpha \to 0}{=} 0,\label{eq:model-solution}
\end{equation}
and amplitude $A$ (For a derivation and the solution for $A$ see Appendix~\ref{app:amplitude}). In the case of the pure displacive model ($\alpha\to 0$), a cosine motion with no additional phase is expected.

The expected phase for various values of $\alpha$, based on the calculated carrier relaxation rates obtained by O'Mahony~\textit{et al.} (see dash-dotted line in Fig.~\ref{fig:residual}b) is shown in Fig~\ref{fig:phase}a. Reasonable agreement with the observed phase, within the range of significantly measured phase, especially above $E_F$, is seen for values of $0.6<\alpha<0.9$.
While the minimal phase in the experiment is observed at the Fermi level, the results from DFT calculations~\cite{omahony2020} predict this minimum to occur at energies slightly below the Fermi level, but the extrema at 0\,eV and -0.3\,eV are otherwise well reproduced.
Since the lifetime strongly depends on the density of states, which should be minimal at the Fermi level, this discrepancy could originate from differences in the calculated DOS.
We find the best agreement between calculated and measured phase for a value of $\alpha=0.8$ (black dashes in Fig~\ref{fig:phase}a). 
This indicates that indeed the relaxation of carriers is the driving force of the phonon motion.
However, the measured phase clearly differs from the case that assumes only the relaxed carriers act on the phonon motion ($\alpha=1$).
Indeed $\alpha\to1$ is not a very physical situation, as it would require the optically excited (i.e. not-relaxed) carriers to exact no force on the phonon motion.
As discussed earlier, at the earliest timescales, the carrier temperature becomes ill-defined as the distribution of excited states does not follow a thermal distribution. However, as was also seen in the component analysis and TTM, a part of the carriers already occupy the same states that are occupied after thermalization, as expressed in the scaling factor $r=0.60\pm0.03$). 
In that sense, the obtained value of $\alpha\approx0.8$ could imply the ratio of the occupation of phonon-driving states at optical excitation vs after relaxation.
Following arguments~\cite{omahony2019,murray2015} brought forward for the driving of the E\textsubscript{g} mode (which is not observed in our experimental configuration), this could be the difference between occupation of non fully-symmetric states (which drive the E\textsubscript{g} mode) and their momentum-relaxation to fully-symmetric states (which would drive the A\textsubscript{1g} mode).
If the momentum-relaxation drives the observed phonon phase, this would explain the differences between the spectral dependence of the observed thermalization lifetimes (Fig.\ref{fig:residual}b) and the phase.
A future study targeting the E\textsubscript{g} mode could shine further light on the early-timescale forces, and investigate the differences in phase and oscillation amplitude of the A\textsubscript{1g} mode and the E\textsubscript{g} mode, as previously suggested~\cite{omahony2019}.

\subsection{Signal Onset}
\begin{figure*}[tb]
\includegraphics[width=\textwidth]{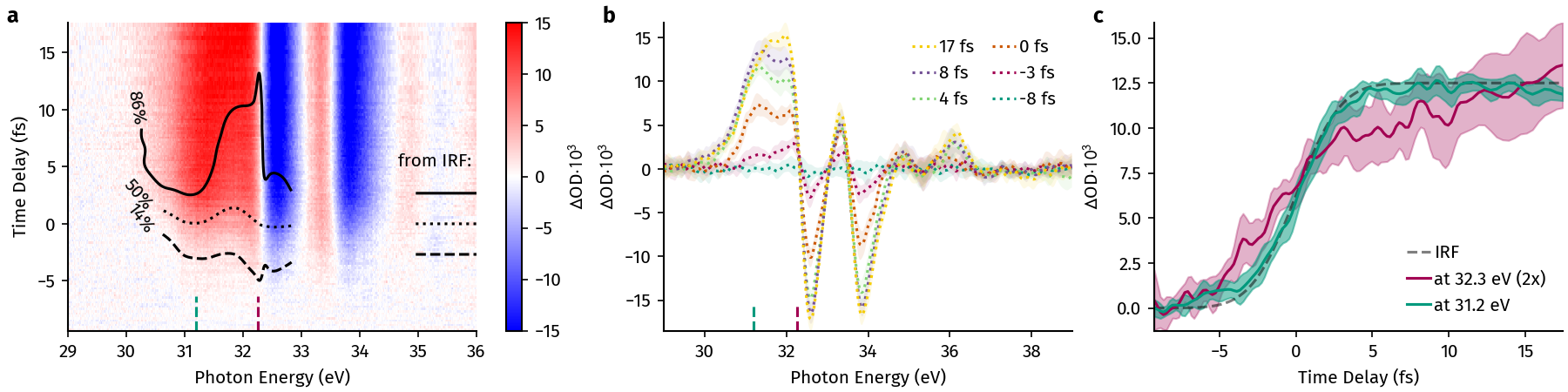}
\caption{\textbf{Transient absorption spectrum during pulse overlap.} \textbf{a} XUV Transient Absorption Spectrum of Sb around the pump pulse. Contour lines show the 14\%, 50\% and 86\% level of the maximum signal at each spectral energy. The inset on the right shows the corresponding levels of the pump fluence from the IRF. \textbf{b} Transient spectra at different delays during the onset of the signal. Strong positive rise in absorbance is observed at energies 0.5\,eV below the Fermi level during the presence of the pump pulse only ($\tau<5$\,fs), while the signal in vicinity of the Fermi level appears to rise early but slower towards later delays. \textbf{c} Temporal dynamic of these different dynamics, showing the deviation from the IRF near the peak of the lattice signal (32.4\,eV), compared to holes in the VB (31.2\,eV, see vertical lines in \textbf{a,b}).
\label{fig:onset}}
\end{figure*}

To analyze the early timescale behavior of the dynamics more carefully, the transient scan is repeated for a short delay range, extending from –9.8\,fs to 17.0\,fs with delay steps of 0.2\,fs (0.4\,fs for the last 6.6\,fs). 

The delay-dependent change in absorbance for the short-time-range scan is shown in Fig.~\ref{fig:onset}a. 
As can be seen, a change in signal quickly appears around zero-delay. However, the signal does not appear to rise uniformly, but rather small, spectrally-dependent deviations from this rise are observed.
To visualize the quantitative differences of the rise signal, contour lines in Fig,~\ref{fig:onset}a signify the 14\%, 50\% and 86\% level of the maximum signal at each spectral energy, with the inset showing the corresponding delay times of the NIR pump fluence. The transient change of absorption at selected delays is shown in Fig.~\ref{fig:onset}b.
Notably, in the valence band, a signal that can be associated with a strong initial hole population around the lower Z- and L-points is observed at photon energies corresponding to 1\,eV to 0.5\,eV below the Fermi level.
The increase in absorption in this region tops out after the pump pulse is over (~5\,fs delay, see Fig.~\ref{fig:onset}b), indicating direct creation of holes in this region due to resonant interband transitions. The absorption in the region closer to the Fermi level, in contrast, keeps increasing and eventually overtakes as the dominant signal, indicating a fast rearrangement of holes into this region.

Compared to the rich dynamics of holes in the valence band, the conduction bands shows a more uniform rise in signal, though again a early peak of the negative change in absorption close to the Fermi level is observed, possibly due to similar reasons as for the holes. The distribution flattens at larger delays and vanishes 1.2\,eV above the Fermi level.

Interestingly, in a small band above the core-to-Fermi-level energy (32.1\,eV) the signal starts to rise towards positive $\Delta$OD at very early delays, but continues to rise at a flatter slope compared to the signal at the lower Z-points or the Gaussian IRF (see Fig.~\ref{fig:onset}c). 
Comparing to the signals obtained from the component analysis (cf. Fig.~\ref{fig:components}) in this region a strong contribution from the lattice dynamic is expected.
It is unclear what the origin of the observed differences in rise-dynamic is.
One possibility could be a different scaling between the signal that originates from carrier occupation (state blocking), which for the conduction band ($>32.1$\,eV) would imply a trend towards negative $\Delta$OD, while the signal from the lattice movement through the shift of the absorption peak would trend towards positive $\Delta$OD.
The implication would be that the start of the lattice movement does not necessarily follow the pump-pulse envelope, which would be consistent with the observation of a carrier-relaxation dependent phase of the phonon oscillation, as observed above on much longer timescales.
Alternatively, a non-linear excitation process could explain the observed temporal behavior. Recent measurements on transient reflectivity in Germanium~\cite{inzani2023} has shown that different non-linear processes can show different dynamics during the envelope of the pump pulse. While no significant transitions were observed at photon energies outside of those accessible by a resonant one-photon transition, due to the narrow band gaps in the vicinity of the Fermi level, a field-induced tunneling process could for example lead to carrier excitation close to the Fermi level, as opposed to resonant one-photon interband excitations. Any non-linear excitation of carriers within the envelope of the pump pulse would not affect the measured phonon phase or decay times significantly, but could lead to an observable different dynamic at the signal onset.
However, a possible explanation for the observed signals at early delays could also be a transient shifting of band-energies during the free induction decay of the core-excited states, i.e. interactions with the pump-pulse during the electronic coherence of the core-excitation. The observed early band would imply the core-to-Fermi-level transition energy to shift towards lower energies due to a NIR induced coupling of bands. Signals from Stark-like shifting is well known from gas-phase~\cite{ott2013,wu2016,drescher2019} and core-exciton~\cite{geneaux2020,lucchini2021,chang2021} attosecond transient absorption spectroscopy, but has so far not been observed for core-to-Fermi-level transitions in (semi-)metals.

Similarly, on the early timescales of the photoexcitation studied here, electronic coherences of the valence-excitation, i.e. coherence of the electron-hole pairs of the NIR-driven interband transitions, could play a role. Such coherences might appear as oscillations in delay at integer multiples of the driving field carrier period~\cite{geneaux2019}. Within the results presented here we do not observe such oscillations. This could hint at a very fast dephasing of the coherences, due to carrier-carrier scattering in the presence of the high carrier density or strongly-coupled carrier-phonon scattering.

The limited signal to noise ratio in the presented data prevents a more thorough analysis, as the signal differences barely surpass the confidence interval of the measurement. An increase in sensitivity in XUV transient absorption spectroscopy, through further stabilization of the light source or referencing technique,\cite{gutberlet2023} would allow to further investigate the onset of the lattice dynamic on the few-femtosecond scale, e.g. by comparing the temporal dynamics seen at higher photon energies that are not overlain by the state-opening/-blocking signals or by studying the sub-cycle behavior of the photoexcitation~\cite{inzani2023}.

\section*{Conclusion}
An XUV transient absorption spectroscopy measurement of the coherent phonon motion in antimony after displacive excitation has been shown. 
The results of the component analysis show the influence of the relaxation of the optically excited and thermally hot carriers on the carrier-dependent signal. 
The results show that the component analysis successfully separates highly transient optically excited carrier dynamics and thermalized carrier and phonon dynamics.
The method obtains equivalent to complementary results as the spectral modelling fit approaches in previous studies~\cite{geneaux2021b,zurch2017} with a soft-modelling approach, i.e. without an a-priori description of the underlying physical dynamics (hard-modelling).
The analysis following an extended temperature model gives meaningful results, and it allows direct comparison to results obtained with other experimental techniques, even though the two temperature model breaks down at the earliest timescales, which reveals the timescale of carrier thermalization. 
The rich spectral information obtained by XUV transient absorption allows to obtain energy dependent relaxation, thermalization and cooling rates of carriers, as well as their effect on the lattice motion.
The obtained spectral phases seem to agree well with theoretical predictions based on the relaxation of carriers, while the measured thermalization time agrees only within the limit of the theoretical predictions, implying that the thermalization is bottlenecked. 
Our results show how XUV transient absorption spectroscopy is able to extract meaningful timescales of electronic and structural dynamics on vastly different scales, ranging from few femtoseconds to picoseconds, from initial scattering rates of carriers, to carrier cooling times, as well as phonon decay and carrier recombination. 
Regarding the DECP model, the spectral dependence of the phonon oscillation phase shows that the phonon motion is not purely displacive, but is dependent on the excited state carrier lifetime to a large degree. In the language of Zeiger~\textit{et al.}~\cite{zeiger1992}, one can conclude that the DECP is driven mainly by the relaxation of carriers and therefore more akin to the change in electron temperature $\Delta T_e$ and not purely the number of excited carriers $\langle n\rangle$. 
This shows that carrier-relaxation needs to be considered to obtain a full picture of dynamics in solids and a pure temperature-based description is not sufficient. 
To optimally control the creation of large amplitude atomic motion through displacive excitation, the effects of the initial relaxation need to be considered.
To further understand the involved dynamics, powerful theoretical methods that couple the ultrafast carrier dynamics as well as the phonon motion are required. Interesting further experimental studies could investigate the efficiency of quenching of the phonon motion by a second NIR pulse, similar as recently done for charge density wave (CDW) materials~\cite{maklar2023} and studying the effect of spectral phase shaping of the pump pulses on the phonon phases.
The role of carrier relaxation is thus expected to be important for many phenomena that underlie manipulation of Peierls distorted materials and optically-driven phase transitions via photoexcitation of carriers, including for example switching of charge-density waves.

\section*{Acknowledgments}

Investigations were supported by the U.S. Air Force Office of Scientific Research, Grants No. FA9550-19-1-0314 (primary) and No. FA9550-20-1-0334, and the W. M. Keck Foundation award No. 046300-002. LD acknowledges the European Union’s Horizon research and innovation programme under the Marie Skłodowska-Curie grant agreement No 101066334 — SR-XTRS-2DLayMat. YSP acknowledges the National Science Foundation Research Experiences for Undergraduates (REU) Grant No. 1852537. 

\section*{Data Availability}
Experimental and computational data that support the findings in this study are available on Zenodo~\cite{drescher_2025_16918641}.

\appendix
\section{Sample Growth}\label{app:sample}
For the experiments, a thin film of Sb is grown on a Si$_3$N$_4$ membrane window (30\,nm thickness, Norcada NX5050X) using a home-built physical vapor deposition oven.
In a vacuum chamber, Sb pellets (99.999\% purity) are mounted on a tungsten coil that is heated to about 300 °C through resistive heating.
Fresh Si$_3$N$_4$ substrates are exposed to the Sb vapor for 6 hours.
A covered part of the edge of the substrate frame during exposure allows to measure the step height of the deposited film using a commercial atomic-force scanning-tip microscope, showing a film thickness of approximately 32\,nm.
Sb is reported to self-crystallize to a polycrystalline form of the rhombohedral A7 phase in thin films~\cite{prins1933} at room temperatures. A Raman spectrum of the films excited by 633\,nm radiation shows the expected prominent phonon peak of the A\textsubscript{1g} mode at 148\,cm$^{-1}$ and E\textsubscript{g} mode at 110\,cm$^{-1}$.

\section{DFT Calculations}\label{app:dft}
Band structure and (projected) density of states were calculated using the Quantum Espresso software package. An ultrasoft pseuopotential with Perdew-Burke-Ernzerhof (PBE) exchange-correlation functional from the GBRV library was used~\cite{garrity2014}. Convergence of the self-consistent field calculations was found at a wavefunction cutoff of 40\,Ry and a grid of 20$^3$ points in k-space in accordance with the standard solid-state pseudopotentials (SSSP) library~\cite{prandini2018}. The lattice was relaxed using the variable cell subroutine, leading to lattice constants of $c=8.694905\,a_0$ and $\cos(\gamma)=0.548392$, and a Peierls parameter $z=0.2328$, in good agreement with crystallographic parameters.

\section{Experimental Setup}\label{app:setup}
To perform XUV transient absorption experiments, NIR pulses centered around 780\,nm wavelength from a commercial Ti:sapphire laser amplifier (Coherent Legend Elite USX operated at 1 kHz repetition rate, 22 fs pulse duration (FWHM) and 4 mJ pulse energy) are spectrally broadened by self-phase modulation in a gas-filled stretched hollow-core fiber ($\approx$700\,µm inner diameter, 2\,m length) filled with a pressure-gradient of argon.
The resulting spectrum spans over one octave.
The pulses are compressed in time using a combination of double-angle dispersion compensating mirrors (Ultrafast Innovations PC70 and PC1332) and various dispersive elements (fused silica windows and wedge-pairs, as well as a 2\,mm thick ADP plate to compensate the negative third order due to self-steepening), leading to pulses shorter than 4\,fs in duration (FWHM) and 1.5\,mJ pulse energy.
A 20:80 beamsplitter is used to divide the pulse energy.
The stronger transmitted pulse is focused by a spherical mirror (f=0.45\,m) and steered through a thin fused-silica window inside a vacuum system. At the focal point a ceramic gas cell with pre-drilled beam openings is continuously flushed with krypton (~23\,Torr) to generate spectrally continuous XUV pulses through the process of high harmonic generation (HHG). Note that the carrier-to-envelope phase was not stabilized during this experiment. The generated XUV pulses are therefore expected to form a short attosecond pulse train instead of an isolated attosecond pulse. A typical XUV spectrum from HHG is shown in Fig.~\ref{fig:hhg}a.
The residual NIR radiation is removed from the XUV by a 200\,nm thick Al filter and the focal point is reimaged by a grazing-incidence toroidal mirror in a 2f:2f configuration. At the second focal point, a motorized XY-stage carrying the sample materials, clean reference membranes and a gas cell is used to carry out the various aspects of the absorption experiments in-situ.

To perform pump-probe time-resolved measurements, the weaker reflected NIR pulses are attenuated by an iris and run parallel to the in-vacuum beam through air.
They are focused by a spherical mirror (f=1.0\,m) and steered through a window into the vacuum system, where they are colinearly recombined by a hole-mirror at 45$^\circ$.
The hole in the center of the mirror allows the XUV probe radiation to pass through, while a part of the NIR excitation pump pulses is reflected from the metal-coated annular surface, leading to a colinear beam geometry.
The pump intensity is limited with an iris aperture to 6\,µJ, leading to an estimated incident fluence of 4.4\,mJ/cm$^2$ at the central ring of the annular NIR focus (170\,$\mu$m beam diameter), assuming an average reflectivity of 70\% across the pump spectrum.
After the interaction with the samples under normal incidence at the focal point, the NIR pump pulses are once more removed from the XUV pulse by an iris acting as a pin-hole and a second aluminum foil (150\,nm thickness).
The XUV pulses are then spectrally dispersed by a flat-field grating (Hitachi 001-0639) and imaged by a CCD camera (Princeton Scientific Pixis 400B) at the image plane.

To control the temporal delay between XUV and NIR pulses, a retroreflector is mounted atop two motorized stages in the pump arm, a long travel stepper-motor controlled stage and a short travel piezo-electric stage.
While the stepper-motor controlled stage allows to scan picosecond ranges, the piezo-electric stage allows finer delay control on the sub-femtosecond scale.
To avoid artifacts from variances in the XUV spectrum, static XUV spectra are rapidly taken sequentially for the sample and blank membranes.
Similarly for pump-probe spectra, a mechanical shutter in the pump arm allows to rapidly take sequential spectra with and without the pump beam present.
The spectra taken without the pump beam are used to calculate a noise correlation matrix from regions of the spectrum where no change of absorbance is observed, and this is used to suppress correlated source noise in the observed transient spectra~\cite{geneaux2019} (see below).

To avoid excessive build-up of heat in the thin Sb sample from the absorption of NIR pulses, a chopper is used to reduce the laser repetition rate to 250\,Hz.
To compensate for drifts due to atmospheric conditions as well as vibrations of the optical components, the pump-probe interferometer can be actively stabilized by a spectral-interference feedback system: By using a small diameter Al film mounted on a larger, annular fused silica plate, part of the 780\,nm HHG driving field is allowed to propagate parallel to the XUV beam.
This annular beam is too large to pass through the center of the hole-mirror and is reflected from the metal surface instead.
There it intersects the part of the pump beam that was not reflected and passed through the mirror instead.
The two focusing beams are overlapped in a spectrometer, leading to spectral interference and a delay-dependent fringe pattern.
The spectral interference over a selected spectral range is read out by a computer and the relative fringe position is used to calculate the deviation of the measured delay from the set-point and used to correct the piezo stage in a proportional-integral controller loop.
While the relative delay is used to stabilize the interferometer during the measurement, the absolute delay between the two pulses is calculated and saved as well and is available for post-processing.
Note that the absolute delay measured in this manner does not directly correspond to the in-situ XUV-NIR delay, due to slight variations in optical path length.
This is in fact deliberate to gain sufficient peak separation of the delay-dependent interference (AC peak) from the structure-rich spectrum of the fiber-broadened pulses (DC peak) in the numerical Fourier evaluation.

\section{Time-Zero and Instrument Response Function Fitting}\label{app:timezero}
\begin{figure*}
    \includegraphics[width=\textwidth]{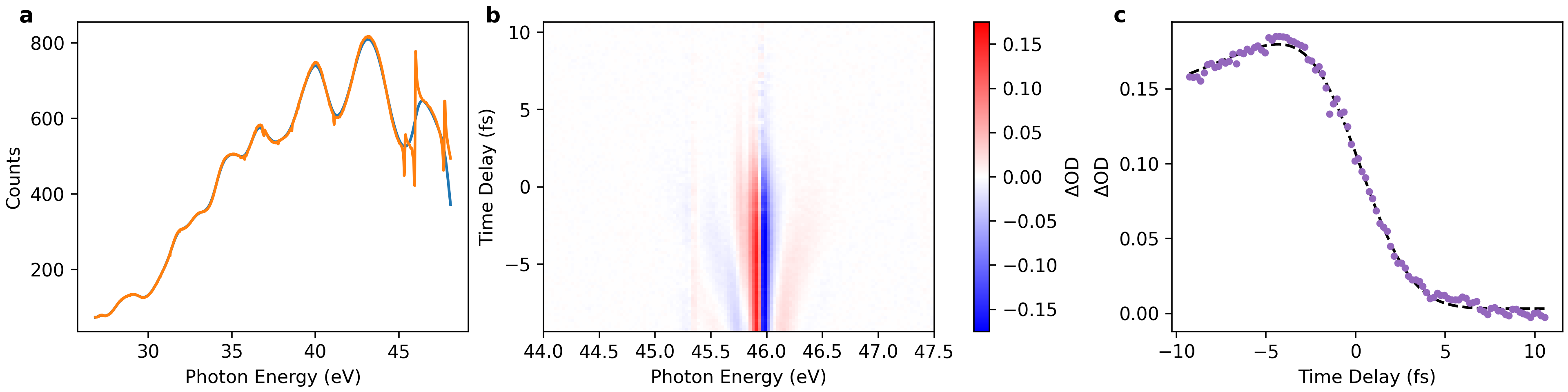}
    \caption{\textbf{Time-Zero and Instrument Response Function Fitting.} \textbf{a} Typical XUV spectrum from HHG before and after absorption by Ne autoionizing resonances. \textbf{b} Transient absorption spectrum of the NIR-induced modification of the Ne 3s4p autoionization resonances. \textbf{c} Temporal dynamic of the NIR-induced modification resembling a exponentially modified error function. A non-linear fit is used to determine zero-delay and the IRF. \label{fig:hhg}}
\end{figure*}
To calibrate the XUV-NIR zero delay, attosecond transient absorption spectra of AC Stark shifting of the Ne 3s4p autoionization resonance are recorded periodically during the delay-dependent scans.
The light-induced phase signal of the autoionization line is analyzed through SVD and the zero-delay is obtained through a non-linear least-squares fitting of a Gaussian instrument response function (IRF) convoluted with the exponential decay of the autoionizing resonance~\cite{drescher2019}.
In difference to the pump-probe spectra recorded for Sb, the Ne spectra where recorded only with the pump present. To suppress the noise originating in a change of the harmonic spectrum, the spectra are high-pass filtered, since the relevant autoionizing lines of Ne are much narrower than the typical harmonic noise~\cite{ott2013}.
Fig.~\ref{fig:hhg}a shows the (averaged) counts of a typical Ne scan (blue) with the low-passed portion of the spectrum (orange) representing the typical HHG spectrum during the experiment.
Fig.~\ref{fig:hhg}b shows the attosecond transient absorption spectrum of Ne after high-pass filtering as a change in absorbance, i.e. after subtracting the static absorption. The temporal component of one exemplary scan is shown in Fig.~\ref{fig:hhg}c with a fitted exponentially modified Gaussian distribution, yielding an IRF of $5.4\pm0.1$\,fs.

\section{Data Processing Steps}\label{app:datapipeline}
\subsection{Data loading, Edge-Referencing and Heat Subtraction}
\begin{figure*}
    \includegraphics[width=\textwidth]{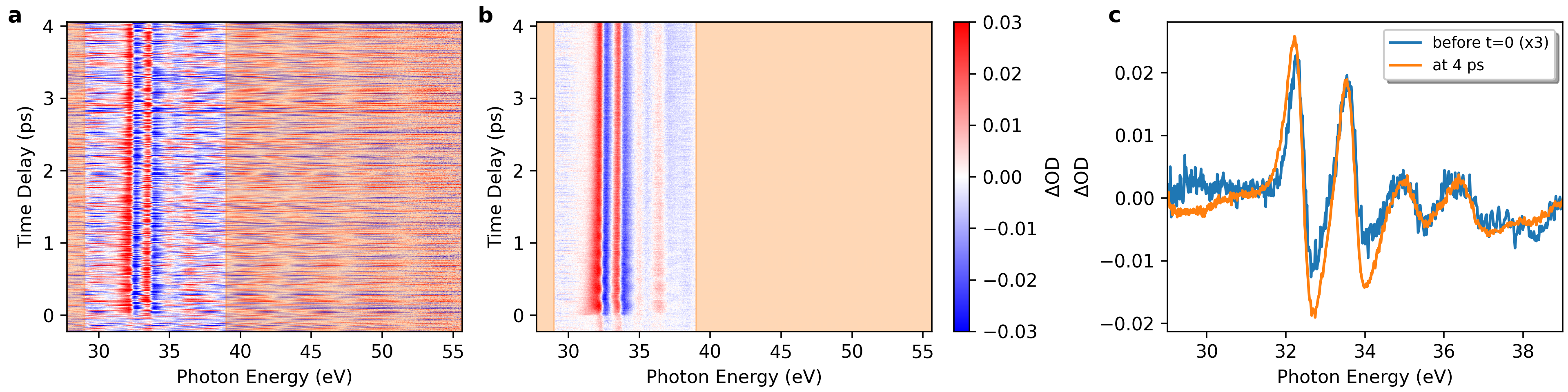}
    \caption{\textbf{Edge-Referencing and Heat Subtraction.} \textbf{a} Raw transient absorption data before edge-referencing in the shaded regions \textbf{b} Result of the edge-referencing. \textbf{c} Comparison of the heat feature before zero-delay to the transient spectrum after 4\,ps. \label{fig:edge_reference}}
\end{figure*}

For the analysis, the spectra taken at each time-delay are loaded into memory sequentially and in two separate arrays for the spectra taken with the pump-beam present and the pump-beam blocked, respectively. Detector dark-count spectra, that were taken before the measurement with the XUV light blocked, are subtracted from each transient spectrum to account for averaged detector dark-noise and potential scattered pump- or ambient light.
For the edge-referencing, we carefully assign areas of no pump-induced change to the Sb absorption spectrum as edge-pixel, which we find to be in the regions above 39eV and below 29eV (shaded regions in Fig.~\ref{fig:edge_reference}(b)).
We note that the results appear robust against variations in the choice of edge-pixel regions.
The edge-referencing is performed simultaneously for all scans of each measurement (i.e. once for the long-delay scan and once for the short-delay scan).
Optical density (OD) is calculated as the (base-10) logarithm of the negative ratio of XUV intensity after propagation through the sample over the intensity before. Similarly, the change in optical density ($\Delta$OD) is the logarithm of XUV intensity with the pump present over without the pump pulse:
\begin{equation}
	\Delta OD(E,\tau) = -\log\frac{I(E,\tau)_\textrm{pump on}}{I(E,\tau)_\textrm{pump off}}
\end{equation}

Fig.~\ref{fig:edge_reference}(a) shows the average change of absorption per delay before edge-referencing and Fig.~\ref{fig:edge_reference}(b) afterwards.
Before zero-delay, a small residual signal can be seen.
The signal, which is interpreted as residual heat lingering between consecutive pump-pulses, is averaged for the first 20 delay points of the scan ($\tau<-80$\,fs) and subtracted from all transient absorption spectra.
Fig.~\ref{fig:edge_reference}(c) shows the subtracted signal.
It is comparable to the signal at the end of the long delay scan ($\tau > 3.9$\,ps), albeit roughly 3x smaller in intensity, supporting our interpretation as residual heat.
Previous investigation of this laser induced heating signal has shown that the removal of the feature does not significantly affect the measured dynamics beyond the effect of an elevated equilibrium temperature~\cite{volkov2019,zurch2017}.
For further analysis, the spectral axis is transformed from wavelength (nm) to photon energy (eV) by linear interpolation.

\subsection{Spin-Orbit Deconvolution}\label{app:SO}
To recover the spectral response of a single core-subshell, we perform spin-orbit deconvolution by Fourier transform~\cite{zurch2017}:
It is assumed that the measured spectra follow a convolution of each spin-orbit component of the $4d$ core-level, such that the measured spectrum $S(\omega)$:
\begin{equation}
S(\omega) = 6 f(\omega) + 4 f(\omega+\omega_\textrm{SO}),
\end{equation}
where $f(\omega)$ would describe the spectral-response for one core-electron excitation in the $4d$ shell, and $\omega_\textrm{SO}=1.2$\,eV the energy splitting between the $4d_{5/2}$ and $4d_{3/2}$ subshell. This implies that the spectral response from each subshell is identical up to the degeneracy factor.
We represent the convolution in Fourier domain as $h(\theta)=1+\frac{4}{6}e(i\omega_\textrm{SO}\theta)$. To then deconvolute the signal and recover the response from a single subshell, the measured spectra are transformed to Fourier domain along their energy axis using a Fast Fourier Transform (FFT) algorithm, divided by $h(\theta)$ and then back-transformed by inverse FFT.

\subsection{Reconstruction of thermal-mode through component analysis}\label{app:svd}
\begin{figure}[tb]
\includegraphics[width=.5\textwidth]{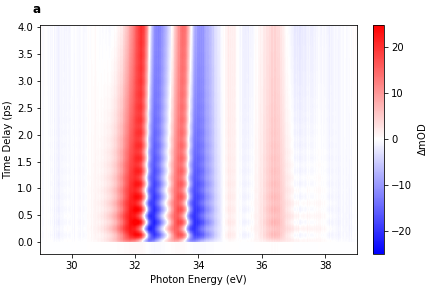}
\includegraphics[width=.5\textwidth]{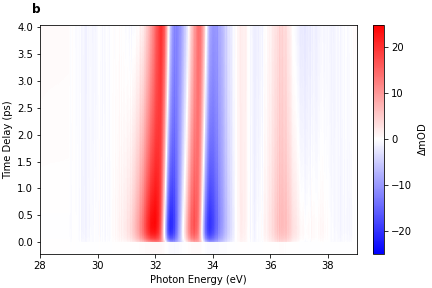}
\caption{Modelled transient absorption spectrograms obtained by combining the spectral components from SVD with analytical temporal dynamics. \textbf{a} Full model from non-linear fit result using Eqn.~(\ref{eq:fit-model}). \textbf{b} Reduced model without coherent phonon oscillations.\label{fig:model}}
\end{figure}
The singular value decomposition applies a bi-linear vector decomposition to the matrix of measured data (spectrogram) $\bm{A}$, where the matrix elements $a_{ij}$ represent the values sampled at time-delay $\tau_i$ and photon-energy $\omega_j$, such that:
\begin{equation}
    a_{ij} = \sum_n s_n u_{ni} v_{jn},
\end{equation}
where $\bm{s}$ is a vector containing the weights of all components, $\bm{u}_n$ and $\bm{v}_n$ the associated temporal and spectral vector for each component, respectively.

To perform the SVD in the basis of the static absorbance, the static absorbance is added back to the measured change in absorbance. The leading component, encoding the static absorbance, is then omitted from the following analysis.
The distribution of weights for the SVD, up to the 20th component, of the long delay-scan is shown in Fig.~\ref{fig:weights}. As expected, the leading strong component of the static dominates, with the two thermal components following.

Having established a model expression for the temporal behavior $f(\tau)$ allows to reconstruct the spectrogram for the model $\bm{A}_\text{model}$ by replacing the temporal vector by sampling the expression at the same delay-points and calculating the outer product with the associated spectral component and scalar weight:
\begin{equation}
    \bm{A}_\text{model} = s_n f(\tau_i) \otimes \bm{v}_n.
\end{equation}

As such, the resulting spectrogram from the non-linear fit of Eqn.~(\ref{eq:fit-model}) below is shown in Fig.~\ref{fig:model}a. Compared to the experimental results shown in Fig.~\ref{fig:static}, there is good agreement with the dominating dynamics. As described previously, the influence of the coherent phonon motion can now be removed from this model, to obtain the pure thermal model dynamic, by reducing the oscillation amplitudes to zero. The resulting spectrogram is shown in Fig.~\ref{fig:model}b. As can be seen the system quickly reaches its maximum after laser-excitation and then starts to equilibriate towards later time-delays. This spectrogram is subtracted from the experimental data to obtain the isolated phonon-dynamics shown in Appendix~\ref{app:fourier}.

\subsection{Confidence Interval Estimation within Singular Value Decomposition}\label{app:CI}
To estimate confidence intervals for the temporal dynamics through our singular value decomposition based analysis, the independent scans are concatenated along the delay axis, such that they share a common spectral axis, prepended by the array of the per-delay average of all scans. To perform the SVD with respect to the static spectrum, the static spectrum is added to all delay-points.
The SVD is then performed for the whole array, meaning that for each spectral component, seven temporal components are obtained, containing the scan-averaged and each independent temporal dynamic associated with the spectral components. For the non-linear least-squares fitting, the function is fitted to the averaged dynamic, while taking the standard deviation for each time-point from the six independent measurements as weights.

\section{Residuals of Component Analysis}\label{app:residuals}
\begin{figure}[tbp]
\includegraphics[width=.5\textwidth]{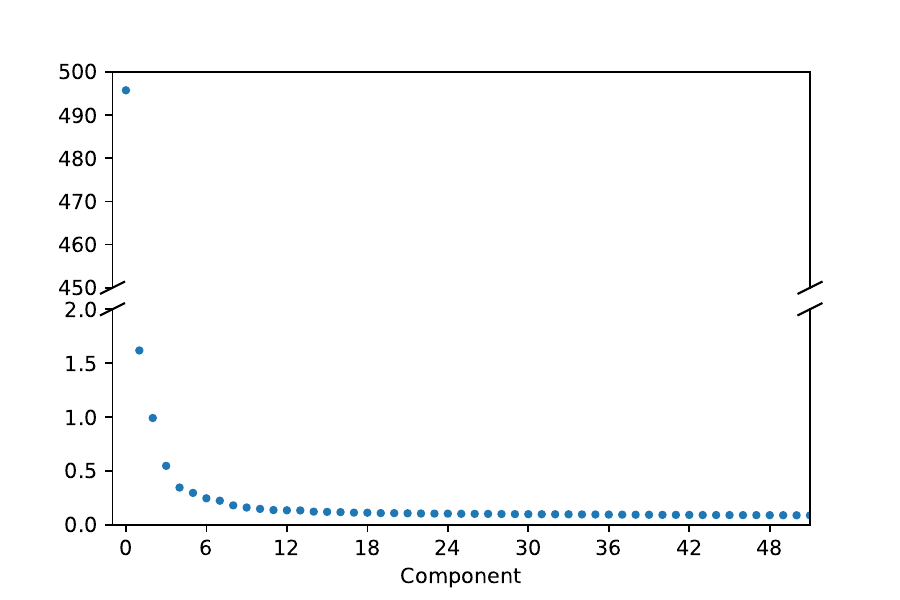}
\caption{\textbf{Weights of the singular value decomposition.}\label{fig:weights}}
\end{figure}

\begin{figure*}[tbp]
\includegraphics[width=\textwidth]{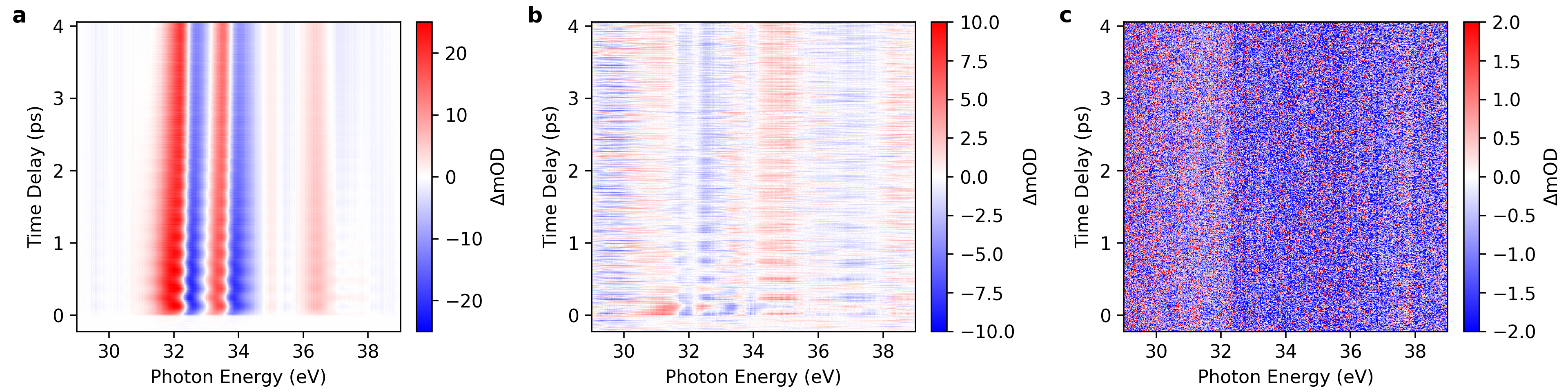}
\caption{\textbf{Residuals of Component Analysis.} \textbf{a}: Transient absorption spectrum of thermal carrier and lattice component. \textbf{b}: Transient absorption spectrum of the following 10 components, showing mostly the non-thermalized carriers at early timescales, as well as residual phonon oscillations that were not captured by either the thermal carrier or lattice component. \textbf{c}: Difference between experimental data and components shown in a and b.\label{fig:residuals}}
\end{figure*}
The distribution of weights from the singular value decomposition (SVD) is shown in Fig~\ref{fig:weights}. As described previously, the added static absorption leads to a strong 0th component weight, which is omitted from the analysis of the change of absorbance. The following two components contain the thermal carrier and lattice dynamics as described above and are shown in Fig.~\ref{fig:residuals}a. The following 10 components are shown in Fig.~\ref{fig:residuals}b. They mostly show the dynamics of non-thermalized carriers, as discussed above, at early timescales. Additionally, residual phonon oscillations are visible throughout the spectrogram, which are not well captured by either the thermal carrier nor the lattice component. This is likely due to the spectral phase dependence of the phonon oscillation, which is not directly describable by a bi-linear decomposition and therefore requires many components of the SVD. Lastly, the transient absorption spectra in Fig.~\ref{fig:residuals}a,b are subtracted from the full experimental data (an equivalent approach would be to calculate the transient absorption spectrum from all other SVD components as the SVD yields identical matrices within the numerical precision of the computer). The result is shown in Fig.~\ref{fig:residuals}c and contains mostly experimental noise with some structural features close to zero delay and at 38\,eV. These are likely due to artifacts in the decomposition of the previously described features and insignificant within the confidence interval (CI) of the experimentally obtained transient absorption data.

\section{Extended Two Temperature Model}\label{app:model}
The extended two temperature model assumes the density of non-thermal carriers $n$:
\begin{equation}
	\dot{n}(t) = P(t)-\rho n(t),\label{eq:eom-carrier_density}
\end{equation}
where $P(t)$ is the pump-pulse intensity and $\rho$ describes the rate of carrier relaxation. The relaxation of non-thermal carriers leads to a change of carrier temperature $T_e$:
\begin{equation}
	C_e \dot{T_e}(t) = \rho n(t)-G_0(T_e(t)-T_l(t))-\frac{F}{V}\dot{Q(t)},\label{eq:eom-carrier_temp}
\end{equation}
where $C_e$ is the carrier heat capacity, $G_0$ the electron-phonon coupling term with the lattice temperature $T_l$. The last term accounts for the back-coupling effect due to the change of potential energy $F=\partial{U}/\partial{Q}$ caused by the phonon motion $\partial{Q}/\partial{t}$ per unit volume $V$~\cite{giret2011}.

The phonon-motion in the DECP model is described as:
\begin{equation}
	\ddot{Q}(t) +\omega_0^2 Q(t)=\frac{F(t)}{m}-2\gamma\dot{Q}(t),\label{eq:eom-phonon}
\end{equation}
where $\omega_0$ is the undampened phonon frequency and $\gamma$ the dampening rate, such that $\omega=\sqrt{\omega_0^2-\gamma^2}$, $F$ the displacement force and $m=2M$ the reduced mass of the phonon ($M$ the atomic mass of Sb).

This leads to the rate equation for the lattice temperature:
\begin{equation}
	C_l \dot{T_l}(t) = G_0(T_e(t)-T_l(t))+\frac{2 m \gamma}{V}\dot{Q}(t)^2
\end{equation}
In the model, it is assumed that the displacement force follows the pump pulse $P$, such that
\begin{equation}
	F(t)\propto	\int_{-\infty}^t P(\tau)d\tau.
\end{equation}

\begin{table}[htb]

    \begin{tabular}{c|c}

        Term & Value \\
        \toprule
         A  & $-82.6\pm0.7$\\
         B  & $-2.5\pm0.8$\\
         C  & $15.6\pm0.5$\\
         D  & $45\pm3$\\
         E  & $23.7\pm0.5$\\
         r  & $0.60\pm0.03$\\
         $\rho$ (ps$^{-1}$) & $8.5\pm0.7$\\
         $\beta$ (ps$^{-1}$) & $0.339\pm0.005$\\
         $\gamma$ (ps$^{-1}$) & $1.04\pm0.04$\\
         $\omega$ (rad THz) & $26.35\pm0.06$\\
         $\varphi$ (rad $\pi$) & $-0.09\pm0.02$

    \end{tabular}
    \caption{Fit results using Eqns.~(\ref{eq:fit-model}).\label{tab:fit}}
\end{table}

Assuming the heat capacities and coupling/relaxation rates described above are constant over the change in temperature and carrier-energy, we can write an approximate general solution in closed form:
\begin{widetext}
\begin{subequations}
	\label{eq:fit-model}
	\begin{equation}
		\Delta T_e(t) \propto P(t-t_0) * \left(A[1-r e^{-\rho t}]e^{-\beta t}+B\cos(\omega t + \varphi)e^{-\gamma t}\right)
	\end{equation}

	\begin{equation}
		\Delta T_l(t) + \Delta Q(t) \propto P(t-t_0) * \left(C(1-e^{-\beta t}) + D(1-e^{-2\gamma t}) + E[1+\cos(\omega t + \varphi)]e^{-2\gamma t}\right),
	\end{equation}
\end{subequations}
\end{widetext}
where $\beta$ is the effective electron cooling rate and the $*$-operator denotes a convolution integral. To account for the instrument response, we assume the pump pulse follows a Gaussian distribution with width 5.4\,fs (FWHM), as obtained from a measurement of the modification of Ne autoionizing lines as described in appendix~\ref{app:setup}.
Note that we find it necessary to introduce the additional scaling parameter $r$ in the equation above, to account for the partial occupation of thermalized states within the pump-pulse duration and the observed slow relaxation (see discussion above).

A multi-dimensional least-squares fit is performed for both channels, where the values for the rates and oscillation frequency are minimized at the same time, while the amplitudes are left as free parameters.
The best-fit prediction is shown as solid lines in Fig.~3 and the values together with a confidence interval estimated from the parameter covariance are shown in Table~\ref{tab:fit}.

\section{Bi-Exponential Model}\label{app:biexponential}
The fit-function used in the bi-exponential fit can be seen as a reduced version of the two-temperature model equation to avoid overfitting the experimentally obtained data. The function used was:
\begin{equation}
    f(t) = P(t-t_0) * \left[ C(1-e^{-\rho t})Ae^{-\beta t}+B\cos(\omega t + \varphi)e^{-\gamma t}\right],
\end{equation}
where $\gamma$ and $\omega$ were fixed to the values obtained from the component analysis.

\section{Amplitude of Coherent Phonon Mode with Carrier-Relaxation}\label{app:amplitude}
We will start with the general solution of the differential equation (cf. Eqn.~(\ref{eq:model-pdq},\ref{eq:model})). Again, following the work by O'Mahony~\textit{et al.}\cite{omahony2019}, we solve using the initial conditions $Q(0)=0$ and $\dot{Q}(0)=0$, which gives us:
\begin{subequations}
\begin{equation}
 	A\cos(\varphi)=\frac{\alpha}{\rho^2+\omega^2}-\frac{1}{\omega^2}
 \end{equation}
 \begin{equation}
 	A\sin(\varphi)=\frac{\alpha}{\rho^2+\omega^2}\frac{\rho}{\omega}.
 \end{equation}
\end{subequations}
 Taking the ratio between these two equations gives Eqn.~(\ref{eq:model-solution}. On the other hand, taking the sum of squares:
 \begin{equation}
 	A^2 = \frac{\rho^2+(\alpha-1)^2\omega^2}{\omega^4 (\rho^2 + \omega^2)}
 \end{equation}
Then
\begin{equation}
	A = \frac{\Lambda}{\omega^2},
\end{equation}
where
\begin{equation}
	\Lambda^2 = \frac{\rho^2+(1-\alpha)^2\omega^2}{\rho^2 + \omega^2} \stackrel{\alpha \to 0}{=} 1.
\end{equation}
We can reinsert the amplitude into Eqn.~(\ref{eq:model}) and get:
\begin{equation}
Q(t) = \frac{\Lambda}{\omega^2}\left(\frac{1}{\Lambda}+\cos(\omega t+\varphi)\right)-\frac{\alpha e^{-\rho t}}{\rho^2+\omega^2}.
\end{equation}

This result is very similar to that reported in Ref.~\cite{omahony2019}, i.e. it represents a lowering of the phonon amplitude due to the time-dependent force and is dependent on the scaling parameter $\alpha$. This implies that the results that were obtained in Ref.~\cite{omahony2019} assuming that a decrease in force on the A\textsubscript{1g}-mode during carrier relaxation could be compatible with the assumed increase in force in this work.

\section{Spectral Phase of Phonon Motion from Fourier Analysis}\label{app:fourier}
\begin{figure}[tbp]
\includegraphics[width=.5\textwidth]{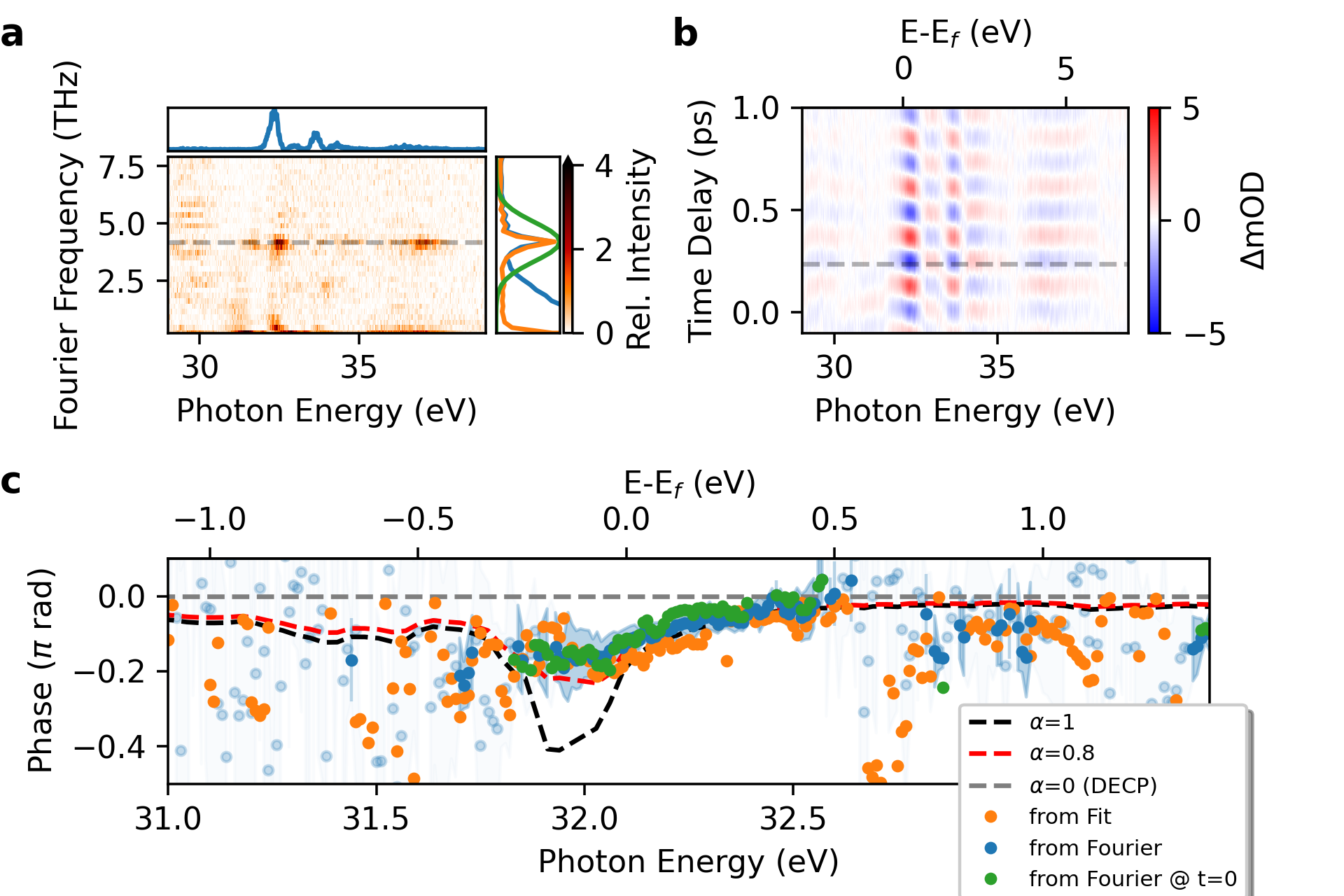}
\caption{\textbf{Spectral Phase of Phonon Motion.} \textbf{a}: Fourier spectrogram of the residual signal (left). The spectral Fourier intensity at the 4.2\,Thz peak is shown on top and shows strong peaks in the vicinity of $E_F$ for both spin-orbit split core-level transitions.  The spectral-integrated Fourier intensity is shown on the right. The isolation of the non-thermalized contributions leads to a high-contrast at the phonon peak (orange) compared to the full signal (blue). A Gaussian filter centered on the phonon frequency is applied for further analysis (green). \textbf{b}: Action of the coherent phonon motion obtained from inversion of the Fourier transform after applying the band-pass filter. A spectral-dependent phase of the oscillation is visible in the vicinity of the core-to-CB transitions. \textbf{c}: Analysis of the spectral phase dependence in vicinity of $E_F$ of the $4d_{5/2}$ core-level. The CI is shown as shaded areas and estimated as described above. Data points with a CI$>\pm\pi/8$ are shown with a reduced opacity. The phase can be reproduced by including lifetime-dependent carrier relaxation into the DECP model (cf. Eqn.~(\ref{eq:model-solution})), dashed lines for different values of $\alpha$ see Eqn.~(\ref{eq:model-pdq}). The phases form the Fourier analysis is in good agreement with the phases found from the non-linear fit presented in Fog.~\ref{fig:phase}.\label{fig:phase-appendix}}
\end{figure}

As an alternative approach to evaluate the phase of the phonon motion, a Fourier analysis using a Takeda-algorithm is performed~\cite{takeda1982,drescher2022}.
The frequency dependent amplitude for each photon energy is calculated by a discrete Fourier transform.
Due to the strong exponential rise- and decay- dynamics of the carrier relaxation, a strong background tail extending out from the DC-component is observed, limiting the contrast at the phonon peak.
The contrast can be severely enhanced and the background eliminated by subtracting the purely thermal model obtained through the component analysis from the measured data.
The resulting Fourier spectogram is shown in Fig.~\ref{fig:phase-appendix}a. The phonon motion is sharply peaked around $f=4.2$\,THz, in good agreement with the result of the fit (see Fig.~\ref{fig:phase-appendix}a).

To obtain the spectral phase of the oscillation, a Gaussian filter of 2\,THz width (FWHM) is applied around the 4.2\,THz peak (at positive frequencies only) and the inverse transformation into the time domain is calculated.

The filtered oscillations are shown in Fig.~\ref{fig:phase-appendix}b and the phase is shown in Fig.~\ref{fig:phase-appendix}c, evaluated at one full period (t=1/f) of the phonon oscillation (dashed line in in Fig.~\ref{fig:phase-appendix}b). The phases show a similar energy-dependence according to the model equations as described above as the result from the non-linear fitting procedure (orange) and are in agreement with each other within their CI.

\begin{figure}[tbp]
\includegraphics[width=0.5\textwidth]{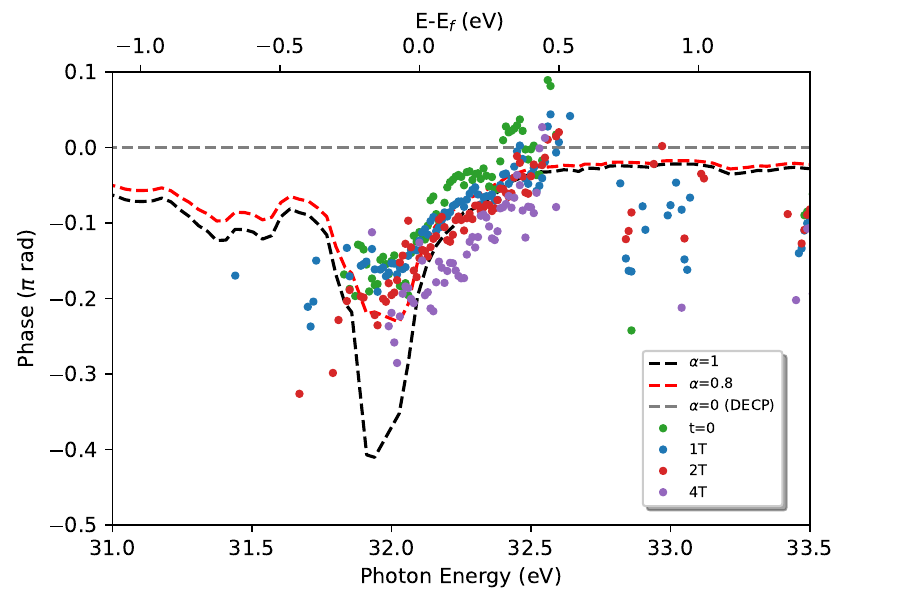}
\caption{\textbf{Phase of Phonon Motion at Different Delays} Phases are analyzed at different full-periods of the phonon oscillation, using the approach presented in Fig.~\ref{fig:phase}.\label{fig:phase-delay}}
\end{figure}

Evaluating the phases at different time delays (see Fig.~\ref{fig:phase-delay}) shows little to no significant variance between different temporal delays. Evaluating the phases after four full oscillations shows the biggest divergence, however this seems to mainly manifest as an overall offset, which could indicate a slight change in phonon frequency due to the carrier cooling and lattice heating.
We note that due to the elimination of the background the results are robust against variations of the filter bandwidth.

\begin{figure}[tbp]
\includegraphics[width=.5\textwidth]{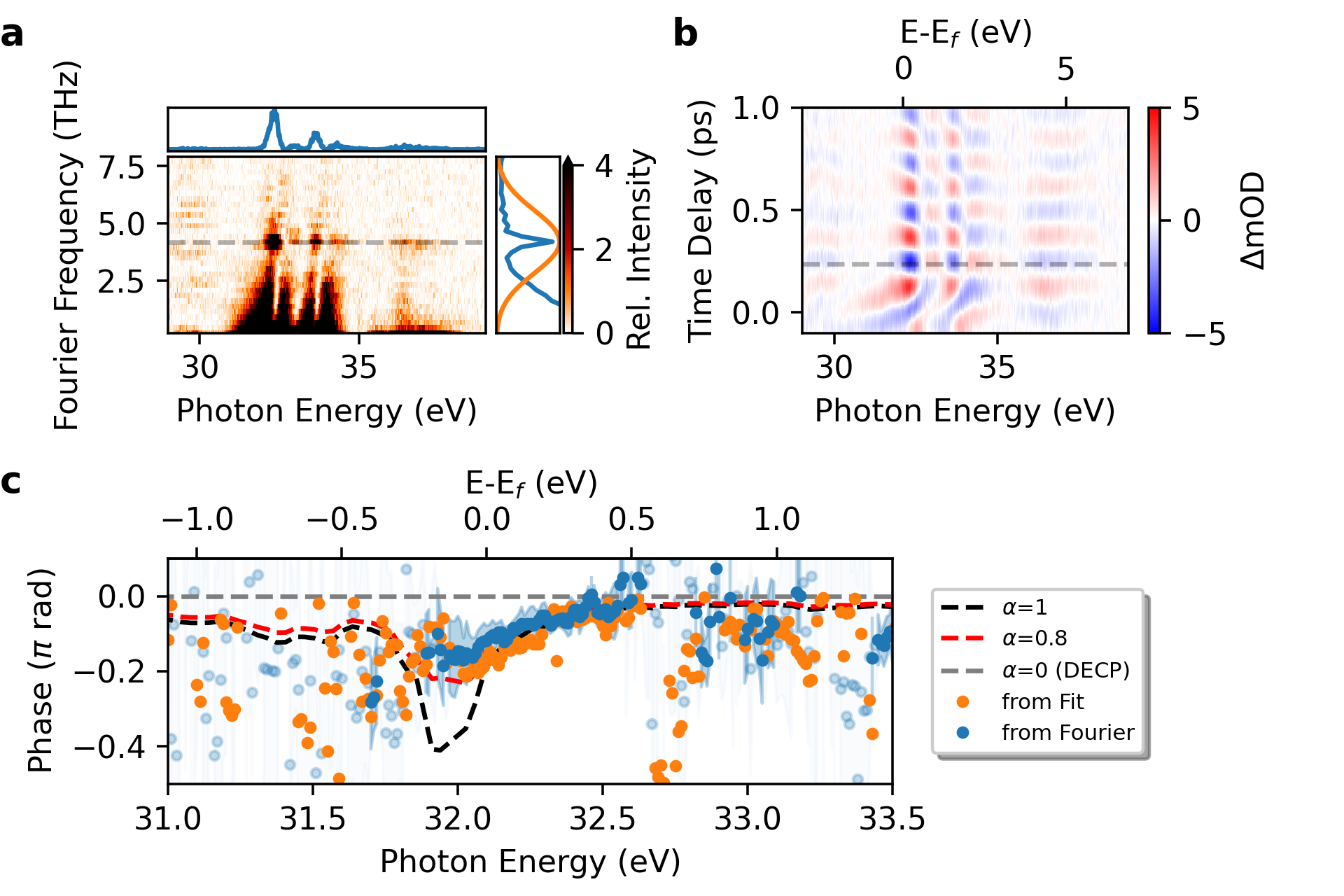}
\caption{\textbf{Spectral Phase of Phonon Motion without Model Subtraction (broad bandwidth).} \textbf{a}: Fourier spectrogram of the full experimental data. \textbf{b}: Action of the coherent phonon motion obtained from inversion of the Fourier transform after applying the broad-band-pass filter. The exponential dynamics at early timescales heavily influence the oscillatory signal before thermalization has occured. \textbf{c}: Analysis of the spectral phase dependence in vicinity of $E_F$ of the $4d_{5/2}$ core-level. \label{fig:phase2}}
\end{figure}

As an alternative approach the model subtraction can be skipped, however in this scenario we find a strong dependence of the bandpass filter width on the phases. This can be seen in Fig.~\ref{fig:phase2}, where the same evaluation as above is performed, although here the experimental data is directly transformed into Fourier domain. In contrast to Fig.~\ref{fig:phase}a, the influence of the exponential dynamics extending from the DC peak can be seen in the Fourier spectrogram. A wide bandpass width (3.2\,THz) is chosen. Due to the strong dynamics of the carrier thermalization, no phonon oscillation is visible at $t=0$ in the reconstructed delay-space spectrogram in Fig.~\ref{fig:phase2}b. However, we still find the phonon oscillation at later delays and evaluation of the phases after one period ($t=1/f$) shows equivalent results as the model subtraction method above (Fig.~\ref{fig:phase2}c).

\begin{figure}[tbp]
\includegraphics[width=.5\textwidth]{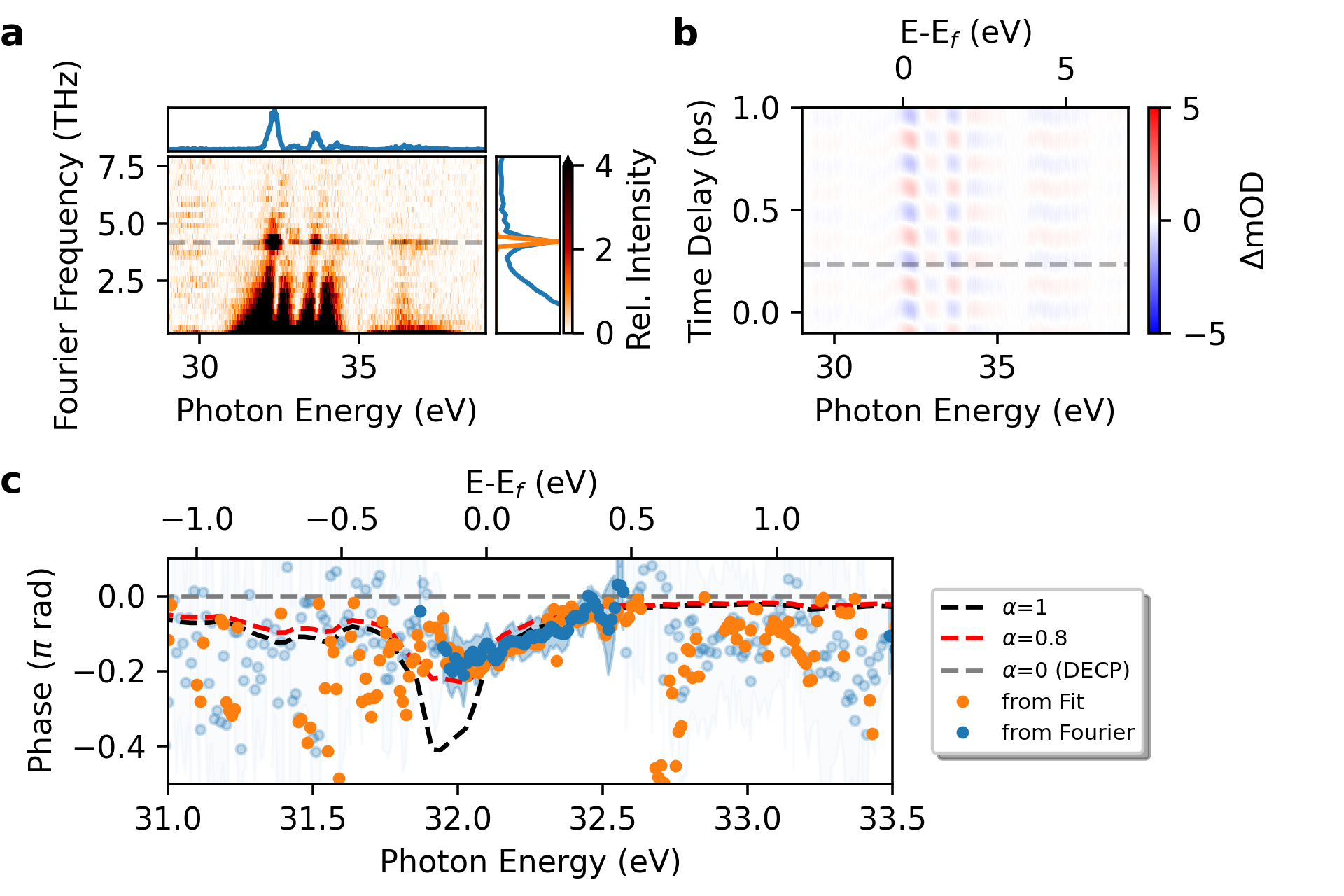}
\caption{\textbf{Spectral Phase of Phonon Motion without Model Subtraction (narrow bandwidth).} \textbf{a}: Fourier spectrogram of the full experimental data. \textbf{b}: Action of the coherent phonon motion obtained from inversion of the Fourier transform after applying the narrow band-pass filter. The exponential dynamics no longer influence the oscillations, which appear very uniform across the delay range. \textbf{c}: Analysis of the spectral phase dependence in vicinity of $E_F$ of the $4d_{5/2}$ core-level. \label{fig:phase3}}
\end{figure}

The wide bandpass filter here allows the oscillatory signal to sufficiently relax such that the influence of the early-timescale dynamics do not significantly affect the phonon signal at later delays.
The opposite can be achieved by choosing a very narrow bandpass width (0.2\,THz), as shown in Fig.~\ref{fig:phase3}. Here, the oscillations can only relax very slowly, as can be seen in the reconstructed delay-dependent spectrogram, which shows little variance across the delay range. Hence, the variation at early delays is suppressed due to the strong, regular oscillations at later delays. We note that the resulting phases agree very well with those obtained from the non-linear fit, which implies that in the fitting procedure the oscillations at later delays strongly influence the result.
Lastly, a similar result can also be achieved by conventional Fourier analysis (i.e. evaluation of the phase directly at the central phonon frequency, without the Takeda algorithm), by cutting the early timescale dynamics ($t<1/f$) from the analyzed transient spectra (not shown).

\bibliography{main.bib}

\end{document}